\documentclass[usenatbib]{mnras}

\setcitestyle{authoryear,round}
\usepackage[T1]{fontenc}
\usepackage{aecompl}

\usepackage{newtxtext,newtxmath}
\usepackage{mdframed,bm}
\usepackage[T1]{fontenc}
\usepackage{ae,aecompl}
\usepackage{graphicx}	
\usepackage{subcaption}
\usepackage{amsmath}	
\usepackage{booktabs}
\usepackage{color}
\usepackage{enumitem}

\usepackage{multirow}
\usepackage[pdftex,dvipsnames]{xcolor}

\usepackage{hyperref}

\hypersetup{
    urlcolor=black,
    citecolor=blue,
    linkcolor=blue,
  }%

\usepackage{physics}
\usepackage{cleveref}
\usepackage{comment}
\usepackage{array}
\usepackage{slashed}

\usepackage{lineno}
\usepackage[dvipsnames]{xcolor}

\usepackage{xspace}
\usepackage{xcolor}
\usepackage{float}
\restylefloat{table}

\usepackage{physics}

\definecolor{bostonuniversityred}{rgb}{0.8, 0.0, 0.0}

\usepackage{eso-pic}

\AddToShipoutPictureBG*{%
  \AtPageUpperLeft{%
    \hspace{0.75\paperwidth}%
    \raisebox{-5.\baselineskip}{%
      \makebox[0pt][l]{\textnormal{DES-2020-0739}}
}}}%

\AddToShipoutPictureBG*{%
  \AtPageUpperLeft{%
    \hspace{0.75\paperwidth}%
    \raisebox{-6.\baselineskip}{%
      \makebox[0pt][l]{\textnormal{FERMILAB-PUB-22-909-PPD}}
}}}%

\author[Prat, Zacharegkas, Park et al.]{
\parbox{\textwidth}{
{\fontsize{11.3}{12} \selectfont 
J.~Prat$^{1,2}$\thanks{E-mail: jprat@uchicago.edu},
G.~Zacharegkas$^{2}$\thanks{E-mail: gzacharegkas@uchicago.edu},
Y.~Park,$^{3}$
N.~MacCrann,$^{4}$
E.~R.~Switzer,$^{5}$
S.~Pandey,$^{6}$
C.~Chang,$^{1,2}$
J.~Blazek,$^{7}$
R.~Miquel,$^{8,9}$
A.~Alarcon,$^{10}$
% ...
\begin{center} (The DES Collaboration) \end{center}
}}
\vspace{0.4cm}
\\
\parbox{\textwidth}{Author affiliations are listed at the end of the paper}}

\begin{document}

\label{firstpage}

\title{Non-local contribution from small scales in galaxy-galaxy lensing: Comparison of mitigation schemes}


\maketitle

\begin{abstract}
  Recent cosmological analyses with large-scale structure and weak lensing measurements, usually referred to as 3$\times$2pt, had to discard a lot of signal-to-noise from small scales due to our inability to accurately model non-linearities and baryonic effects. Galaxy-galaxy lensing, or the position-shear correlation between lens and source galaxies, is one of the three two-point correlation functions that are included in such analyses, usually estimated with the mean tangential shear. However, tangential shear measurements at a given angular scale $\theta$ or physical scale $R$ carry information from all scales below that, forcing the scale cuts applied in real data to be significantly larger than the scale at which theoretical uncertainties become problematic. Recently there have been a few independent efforts that aim to mitigate the non-locality of the galaxy-galaxy lensing signal. Here we perform a comparison of the different methods, including the Y-transformation, the Point-Mass marginalization methodology and the Annular Differential Surface Density statistic. We do the comparison at the cosmological constraints level in a combined galaxy clustering and galaxy-galaxy lensing analysis. We find that all the estimators yield equivalent cosmological results assuming a  simulated Rubin Observatory Legacy Survey of Space and Time (LSST) Year 1 like setup and also when applied to DES Y3 data. With the LSST Y1 setup, we find that the mitigation schemes yield $\sim$1.3 times more constraining $S_8$ results than  applying  larger scale cuts without using any mitigation scheme.
\end{abstract}




\section{Introduction}
\label{sec:intro}
When the light of background (\textit{source}) galaxies passes close to foreground (\textit{lens} or \textit{tracer}) galaxies it gets perturbed, distorting the image of the source galaxies we observe. Galaxy-galaxy lensing refers to the cross-correlation between source galaxy shapes and lens galaxy positions. The amount of distortion is correlated with the properties of the lens sample and the underlying large scale structure it traces. In the case of a spherical distribution of matter, the shear at any point will be oriented tangentially to the direction toward the center of symmetry. Then, the tangential component of the shear will capture all the gravitational lensing signal produced by a spherically symmetric distribution of mass. Because of this, the \textit{tangential shear} averaged over many source-lens galaxy pairs is one of the typical measurements that is done to detect this correlation.

Galaxy-galaxy lensing has had a wide range of applications since it was first detected in \citet{Brainerd_1996}. It has been extensively used to probe the galaxy-matter connection at small scales, e.g. \citet{Choi2012}, \citet{Yoo_2012}, \citet{Kuijken2015}, \citet{Clampitt_2016}, \citet*{Park_2016}, \citet{Zacharegkas2022} or \citet{Luo2022}; to extract cosmological information using the well-understood large scales in combination with other probes such as galaxy clustering and/or CMB lensing as in \citet{Mandelbaum_2013}, \citet{Baxter_2016},  \citet{Joudaki_2017}, \citet{van_Uitert_2018}, \citet*{Y1GGL},  \citet{Singh_2019}, \citet{Lee2022}, \citet{Prat2022}; to obtain lensing shear geometric constraints e.g. \citet{Jain2003},  \citet*{Prat2019}, \citet{Hildebrandt_2020}, \citet{Giblin_2020}, \citet*{desy3-shearratio}, and also recently in \citet{Leauthaud2022} to assess the consistency of lensing across different data-sets and to carry out end-to-end tests of systematic errors.

Moreover, galaxy-galaxy lensing is a standard part of the so-called 3$\times$2pt analyses that combine large-scale structure and weak lensing measurements to extract cosmological information. The 3$\times$2pt stands for the combination of three two-point correlation functions: the autocorrelation of the positions of galaxies (galaxy clustering), the cross-correlation of galaxy shapes and galaxy positions (galaxy-galaxy lensing) and the autocorrelation of galaxy shapes (cosmic shear). This combination was originally proposed in \citet{Hu2004}, followed by \citet{Bernstein_2009} and \citet{Joachimi2010} and has since then been applied to different galaxy survey data sets, such as to KiDS data, as in \citet{Heymans2021}, and to DES data, e.g. in \citet{y3-3x2pt}.

These kind of analyses commonly use the well-understood large scales, placing stringent scale cuts to remove the parts of the data vector that currently add too much uncertainty in the model due to non-linearities in the matter power spectrum, galaxy bias and baryonic effects, amongst others (e.g. \citealt{y3-methods}). However, the galaxy-galaxy lensing signal is non-local in nature,  that is, the predicted signal at a given separation between a source and a lens galaxy (at the redshift of the lens galaxy) depends on the modeling of all scales within that separation, including the non-linear small scales. This can be appreciated expressing the tangential shear of a single lens-source galaxy pair separated by an angular distance $\theta$ as a function of the excess surface mass density $\Delta \Sigma$:
\begin{equation}
\gamma_{t} \, (\theta, z_l, z_s) = \frac{\Delta \Sigma \, (\theta)}{\Sigma_{\mathrm{crit}} (z_l, z_s)},
\end{equation}
where $\Delta \Sigma$ can be expanded as the difference between the mean surface mass density \textit{below} a certain angular scale ($<\theta$) and the surface mass density \textit{at} this given scale $\theta$:
\begin{equation}\label{eq:deltasigma-intro}
\Delta \Sigma\,  (\theta) = \overline{\Sigma} \, (<\theta) - \Sigma\,  (\theta),
\end{equation}
where the non-locality of the tangential shear quantity becomes apparent, since the tangential shear defined at some $\theta$ value will always carry information of all the scales below this value. $\Sigma_\mathrm{crit}$ is just a geometrical factor that depends on the angular diameter distances to the lens galaxy $D_{l}$, the one between the lens and the source $D_{ls}$ and the one to the source galaxy $D_s$, and is defined as:
\begin{equation}\label{eq:inverse_sigma_crit}
\Sigma_{\mathrm{crit}}^{-1} (z_{\rm l}, z_{\rm s}) = \frac{4\pi G}{c^2} \frac{D_{ls} \, D_l} {D_s} \qquad \mathrm{if} z_s > z_l,
\end{equation}
and zero otherwise. In the equation above $G$ is the gravitational constant and $c$ is the speed of light. 

Therefore, our inability to model the small scales accurately enough given the measurement uncertainties impacts the lensing signal at all scales. Expanding on this, in order to predict the lensing signal, an accurate prediction for the galaxy-matter correlation function $\xi_{\rm{gm}}(r)$ is required for some range of physical (3D) scales $r$, see e.g. \citet{MacCrann2020}. At large scales, we expect linear theory to hold and thus we can relate the galaxy-matter correlation function with the matter correlation through a linear galaxy bias factor: $\xi_{\rm{gm}}(r)=b \xi_{\rm{mm}}$, see \citet{Desjacques2018} for a galaxy bias review. At smaller (nonlinear) scales we do not currently have a precise theory to model $\xi_{\rm{gm}}(r)$. Therefore, it is crucial to ensure that the tangential shear measurement is only sensitive to scales in $\xi_{\rm{gm}}(r)$ where we know that the modelling is sufficiently accurate. Since the galaxy-galaxy signal receives a non-local contribution that depends on scales in  $\xi_{\rm{gm}}(r)$ that are much smaller than the separation at which the measurement is made (i.e. the impact parameter in the lens redshift), this non-locality can then force the scale cuts applied in real data to be significantly larger than the scale at which theoretical uncertainties become problematic. For example, due to this reason, the scale cuts in the DES Y1 3$\times$2pt cosmological analysis \citep{desy1kp-3x2} were higher for the galaxy-galaxy lensing part (12 Mpc/$h$) than for the galaxy clustering part (8 Mpc/$h$).

To help with this issue, there have been a few independent efforts to mitigate the non-locality of the galaxy-galaxy lensing signal. The first that was proposed was the annular differential surface density estimator by \citet{Baldauf2010}, which consists of modifying the data vector on all scales in a way that removes information about the lens mass distribution below a chosen scale $R_0$, based on the measured data vector around $R_0$. Later on \citet{MacCrann2020} proposed a methodology to take into account the non-locality by analytically marginalizing over an enclosed point-mass directly when performing the cosmological parameters inference. Finally another estimator-based methodology was proposed by \citet{Park2021}, which achieves the localization of the galaxy-galaxy lensing signal by performing a linear transformation of the tangential shear quantity. In the recent DES Y3 3$\times$2pt work \citep{y3-3x2pt}, the point-mass marginalization methodology was applied to remove the information from small scales above a certain scale, which resulted in being able to model the galaxy-galaxy lensing until 6 Mpc/$h$, a much smaller scale cut than the one used in DES Y1 of 12 Mpc/$h$, even with smaller measurement uncertainties, while the galaxy clustering scale remained the same as in the Y1 analysis (8 Mpc/$h$).

In this paper we perform a thorough comparison of these different proposals with the main goal of understanding which of them is advantageous to use in future cosmological analyses. First, we use simulated data vectors with uncertainties mimicking the LSST Y1 settings to perform such a comparison  and then apply it to DES Y3 data.

\section{Theory}
\label{sec:theory}
\begin{figure}
\begin{center}
\includegraphics[width=0.49\textwidth]{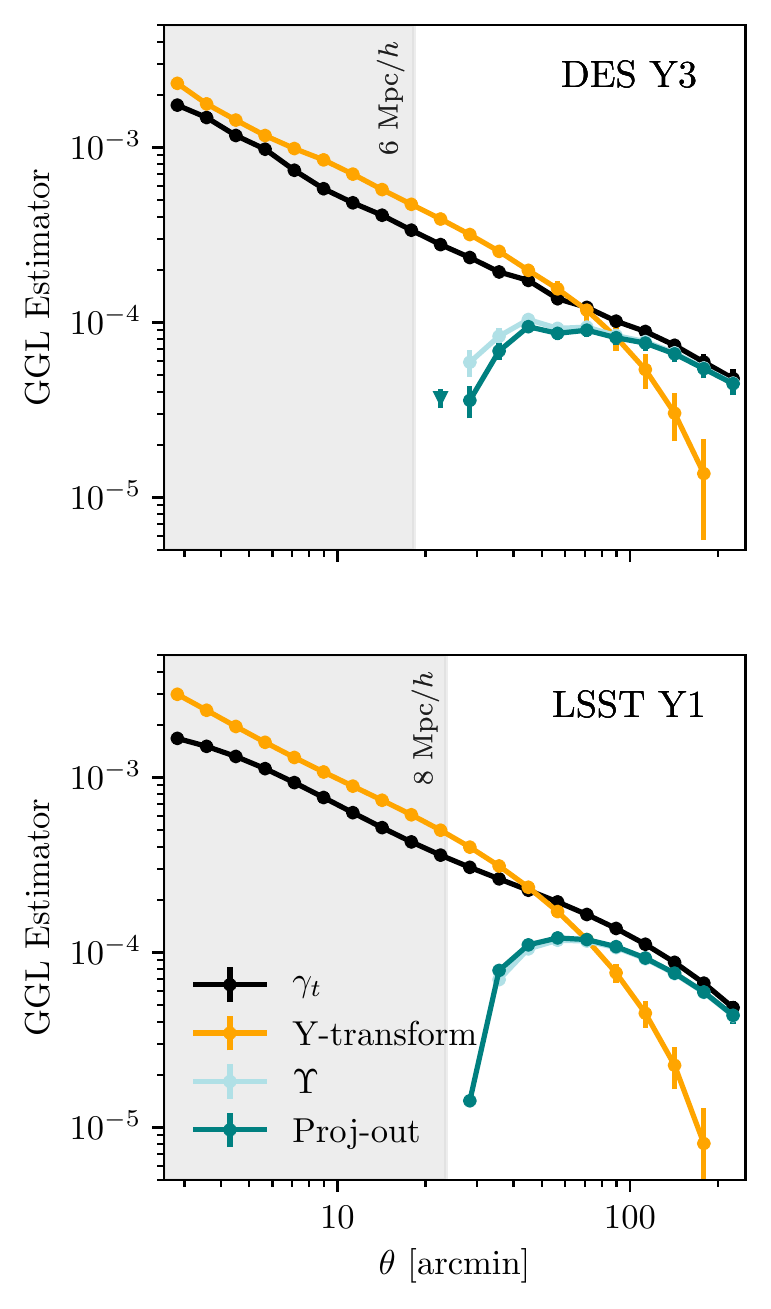}
\end{center}
\caption{Visualization of the data vectors for each of the galaxy-galaxy lensing (GGL) estimators that we compare in this paper to localize the original tangential shear measurements ($\gamma_t$), corresponding to the second lens redshift bin and the highest source bin. We do not add the point-mass marginalization case since it does not involve a modification of the data vector, only of the (inverse) covariance. The triangle point for the proj-out case represents a negative point. }
\label{fig:datavecs}
\end{figure}

Eq.~(\ref{eq:deltasigma-intro}) can be rewritten as a function of physical scale $R=\theta D_l$ in the small angle approximation as:
\begin{equation}\label{eq:delta-sigma-R}
    \Delta \Sigma (R) = \overline{\Sigma} \, (0, R) - \Sigma\,  (R).
\end{equation}
Then, expanding each of the terms, the surface density at a given transverse $R$ scale between the lens galaxy and the source light can be expressed as the integral of the three-dimensional galaxy-matter correlation function $\xi_{gm}(r )$
over the line-of-sight distance $\Pi$, with $r = \sqrt{\Pi ^2 + R^2}$:
\begin{equation}
    \Sigma (R) = \overline{\rho}_m \int_{-\infty}^{\infty} \mathrm{d} \Pi  \left[ 1 + \xi_{gm} \left(\sqrt{\Pi ^2 + R^2} \right) \right] \, , 
\end{equation}
where $\overline{\rho}_m$ is the mean matter density. The other term in Eq.~(\ref{eq:delta-sigma-R}) is the mean surface density between two transverse positions, which can be generally expressed as:
\begin{equation}
    \overline{\Sigma} (R_1, R_2) = \frac{2}{R^2_2 - R^2_1} \int_{R_1}^{R_2} \Sigma (R') R' \mathrm{d} R'. 
\end{equation}
For $R_1 = 0$ and $R_2 = R$ it simplifies to:
\begin{equation}\label{eq:mean-sigma-0}
    \overline{\Sigma} (0, R) = \frac{2}{R^2} \int_0^R \Sigma (R') R' \mathrm{d} R'.
\end{equation}
This term is the one containing the information from all scales down to $R=0$, including the one halo regime for which we do not have an accurate model for $\xi_{gm}$. Assuming we can model $\xi_{gm}$ accurately only down to some minimum scale $r_\mathrm{min}$, we do not want $\Delta{\Sigma} (R)$ to depend on $\xi_{gm}$ below  $r_\mathrm{min}$. From the equation above it becomes clear that $\overline{\Sigma} (0, R) $, and thus $\Delta{\Sigma} (R)$, are only sensitive to the \textit{total} mass enclosed inside $R$ (i.e. to some integral of $\xi_{gm}$), but not to how the mass is distributed (i.e. to the shape of $\xi_{gm}$). For instance, $\Delta{\Sigma}$ at scales larger than $R$ will be the same for a point-mass distribution as for an NFW profile. Also, as shown by Eq.~(\ref{eq:mean-sigma-0}), the contribution from the total enclosed mass will scale as $1/R^2$. This is the key fact that all the estimators described below rely on to remove the dependency of the enclosed mass (or non-locality) of the galaxy-galaxy lensing measurements. Below we summarize each of these currently existing methodologies. We also visualize the modified data vectors for each of the methodologies in Fig.~\ref{fig:datavecs}.

\subsection{Annular Differential Surface Density $\Upsilon$}\label{sec:upsilon}

The Annular Differential Surface Density statistic, $\Upsilon(R)$, is defined in the following way \citep{Baldauf2010}:
\begin{align}
    \Upsilon (R; R_0) & \equiv \Delta \Sigma (R) - \frac{R_0^2}{R^2}\Delta \Sigma (R_0) =\\
    & = \frac{2}{R^2}\int_{R_0}^{R} \mathrm{d} R' R' \Sigma (R') - \frac{1}{R^2}\left[ R^2 \Sigma (R) - R_0^2 \Sigma (R_0) \right] \,. 
\end{align}
From the integration limits it is clear this estimator does not include information from scales below $R_0$, because  $\Delta \Sigma (R_0)$ contains the same small-scale contribution as $\Delta \Sigma (R)$, just rescaled by $R^2/R_0^2$.  The second line follows from the first one by substituting $\Delta \Sigma (R) =  \overline{\Sigma} \, (0,R) - \Sigma\,  (R)$ and using the definition
\begin{equation}
    \overline{\Sigma} (R_1, R_2)= \frac{2}{R_2^2-R_1^2}\int_{R_1}^{R_2} \mathrm{d} R' R' \Sigma (R') . 
\end{equation}
In this estimator and in the ones below that also involve a transformation of the data vector, the model is transformed in the same way as the measurements.
 Moreover, the Annular Differential Surface Density statistic can be equivalently defined for the tangential shear quantity using  angular scales:
\begin{equation}\label{eq:upsilon}
    \Upsilon_{\gamma_t} (\theta; \theta_0) \equiv \gamma_t (\theta) - \frac{\theta_0^2}{\theta^2}\gamma_t (\theta_0) \, ,\\
\end{equation}
using the small-angle approximation to go from $R$ to $\theta$. In realistic scenarios, where each lens tomographic bin has a non-negligible width, a given value $\theta_0$ will mix a range of physical scales $R_0$. In this work we  use the Annular Differential Surface Density statistic based on the tangential shear quantity throughout the paper (instead of the $\Delta \Sigma$ one), and use $\theta_0$ cuts computed with the mean $z_l$ for each redshift bin, given a specified $R_0$ cut. In this paper we use values for $R_0$ of 6 Mpc$/h$ and of 8 Mpc$/h$, depending on the data set and its constraining power.

The covariance of $\Upsilon$ will also generally need to be modified. Given that we can write the transformation as $\Upsilon = \gamma_t - X$, with $X$ being a constant, then $\mathrm{Var}[\Upsilon] = \mathrm{Var}  [\gamma_t] + \mathrm{Var}  [X]  - \mathrm{cov} [\gamma_t, X]$. In the case that $X$ is noiseless, the covariance of $\Upsilon$ will be identical to the $\gamma_t$ one.  In our implementation we always modify the covariance of the $\Upsilon$ statistic to include the noise of $\gamma_t(\theta_0)$.

\subsection{Y-transform}\label{sec:y-transform}

The $Y$-transformation derived in \citet{Park2021} is a localizing linear transformation that utilizes the local quantity $\Sigma(R)$ underlying the galaxy-galaxy lensing observable $\Delta\Sigma(R)$. By inverting the $\Delta\Sigma(R)$--$\Sigma(R)$ relationship, the $Y$ quantity defined as
\begin{align}\label{eq:ycont}
    Y(R) &\equiv \Sigma (R) - \Sigma(R_\text{max}) \nonumber \\
    &= \int^{R_\text{max}}_R\mathrm{d}\ln R' \left[ 2 \Delta \Sigma (R') + \frac{d \Delta \Sigma (R')}{d \ln R'} \right]
\end{align}
recovers the local $\Sigma(R)$ up to an overall additive constant $\Sigma(R_\text{max})$. To treat the discretized observables most frequently used in real analyses, this relation is also discretized to a linear transform given by
\begin{equation}
    \mathbf{Y} = \left(2\mathbf{S} + \mathbf{S}\mathbf{D}\right) \mathbf{\Delta\Sigma} = \mathbf{T}\mathbf{\Delta\Sigma},
\end{equation}
where the trapezoidal summation matrix $\mathbf{S}$ representing the log integral and the finite differences matrix $\mathbf{D}$ representing the log differentiation are used to define the final transformation matrix $\mathbf{T}$. The linear format of this transformation allows a further direct application to a $\gamma_t$ vector, as $\gamma_t$ is proportional to $\Delta\Sigma$. Thus, by transforming an observed galaxy-galaxy lensing vector $\bm{\gamma}_t$ and its covariance $\mathbf{C}_{\gamma}$ as
\begin{eqnarray}
    \mathbf{Y}_\gamma & = & \mathbf{T}\bm{\gamma}_t, \\
    \mathbf{C}_{Y_\gamma} & = & \mathbf{T}\mathbf{C}_\gamma \mathbf{T}^\mathbf{T},
\end{eqnarray}
we achieve a likelihood analysis with a localized galaxy-galaxy lensing observable. It is notable that the $\mathbf{T}$ matrix nulls out components in $\mathbf{\Delta\Sigma}$ ($\bm{\gamma}_t$) proportional to $1/R^2$ ($1/\theta^2$), which can also be seen from Eq.~(\ref{eq:ycont}) when a term proportional to $1/R^2$ is added to $\Delta\Sigma(R)$. Another way to see this is that if $\Delta\Sigma(R)$ has a $1/R^2$ shape, the integral from Eq.~(\ref{eq:ycont}) vanishes. Note that this is also true for the $\Upsilon$ statistic of section~\ref{sec:upsilon}, i.e. adding a term proportional to $1/\theta^2$ makes no difference to the estimator (see Eq.~\eqref{eq:upsilon}).

\subsection{Point-mass marginalization}\label{sec:pm}

\citet{MacCrann2020} proposed to analytically marginalize over the contribution from within the small-scale cut by treating it as a point-mass (PM) contribution scaling as $1/R^2$. This point-mass term can be expressed as an addition to the tangential shear model for a given lens redshift bin $i$ and source redshift bin $j$: 
\begin{equation}\label{eq:point-mass}
\gamma_t^{ij} \, (\theta) = \gamma_{t, \rm model}^{ij} \, (\theta) + \frac{ A^{ij}}{\theta^2}\, ,  
\end{equation}
where we use the small-angle approximation to go from $R$ to $\theta$. Here the $\gamma_{t, \rm model}^{ij}$ is a prediction based on a model for the 3D galaxy-matter correlation function $\xi_{\text{gm}}(r)$ that is correct for scales $r > r_\text{min}$, but can be arbitrarily wrong for $r < r_\text{min}$, and $A^{ij}$ is some unknown constant that we can marginalize over. The simple form of this contamination model makes it suitable for an analytic marginalization approach given that the scale dependence is not dependent on cosmology or the lens galaxy properties. We want to marginalize $P(\gamma_{t, \text{obs}}\, (\theta) |\gamma_{t, \text{model}}\, (\theta) , A)$ over the unknown constant $A$ in order to obtain the likelihood we are interested in, namely:
\begin{equation}
    P(\gamma_{t, \text{obs}}\, (\theta) |\gamma_{t, \text{model}}\, (\theta) )=\int\mathrm{d}A \, P(A) P(\gamma_{t, \text{obs}}\, (\theta) |\gamma_{t, \text{model}}\, (\theta) , A).
\end{equation}
In the case that the $\gamma_{t, \text{model}}$ is Gaussian distributed with covariance matrix $\mathbf{C}_\gamma$, and we have a Gaussian prior on $A$ with mean zero and width $\sigma_A$, one can show that \citep{Bridle2002}
$P(\gamma_{t, \text{obs}}\, (\theta) |\gamma_{t, \text{model}}\, (\theta) )$ is also Gaussian distributed with a covariance matrix
\begin{equation}
    \mathbf{N}= \mathbf{C}_\gamma + \sigma^2_A \Vec{x}\Vec{x}^\text{T},
\end{equation}
where $\Vec{x}$ has elements $x_n = (\theta_\text{min}/\theta_n)^2$, and $\theta_{\text{min}}$ represents the scale cut. This means that in order to marginalize over the free parameter $A$, we only need to add this term to the original covariance rather than explicitly sampling over possible values of $A$ in e.g. an MCMC chain. In this work we use an infinite prior for $\sigma_A$. In this case, the extra term is added to the inverse covariance directly, following the procedure described in \citet{MacCrann2020} and in \citet{Prat2022}. 

\subsubsection{Point-mass marginalization using geometric information} \label{sec:point-mass-geometric}
The amplitudes $A^{ij}$ can be written as
\begin{align}
  A^{ij}  = \int \mathrm{d}z_l \int \mathrm{d}z_s n^i_l (z) n^j_s (z) B^i(z_l) \Sigma^{-1}_{\rm crit}(z_l,z_s) D^{-2}(z_l) \nonumber 
\end{align}
where $B^i$ represents the total mass enclosed within $\theta_\text{min}$ for the $i$-th lens bin, $n^i_l(z)$ is the redshift distribution of each lens bin, $n^j_s(z)$ for each source bin, and $D(z_l)$ is the angular diameter distance to the lens redshift $z_l$. If we assume that 
 $B^i$ evolves slowly across the width of the lens bins we can  drop the $z_l$ dependence and define 
 the parameters $\beta_{ij}$ in the following way:
\begin{equation}
 A^{ij}    \approx B^i \int \mathrm{d}z_l \int\mathrm{d}z_s n^i_l (z) n^j_s (z) \Sigma^{-1}_{\rm crit}(z_l,z_s) D^{-2}(z_l) \nonumber  \equiv B^i \beta_{ij}.
  \end{equation}
The parameters $\beta_{ij}$  are purely geometrical (sometimes also called shear-ratio information), and thus can be exactly computed analytically given the input redshift distributions\footnote{Note that since
 the second term of the RHS of Eq.~(\ref{eq:point-mass})
only  accounts for the unmodeled enclosed mass caused by the mismodeling of the halo-matter correlation function, other effects such as IA or magnification would be  a part of the first term of the RHS of Eq.~(\ref{eq:point-mass}), and thus do not need to be included in the computation of the $\beta_{ij}$ factors.}. Then, the predicted $\beta_{ij}$ factors can be used to reduce freedom in the model by fixing the relative scales between the source bins sharing the same lens bin and only marginalizing over $B^i$ instead of over a free-form $A^{ij}$.

We label this variant of the model as "Point-mass (free per $z^i_l$)", since in this case there is only one free parameter per lens bin. When this approximation is not used (i.e. only using Eq.~\ref{eq:point-mass}) we label the model as "Point-mass (free per $z^i_l-z_s^j$)". In the DES Y3 analysis, where the point-marginalization was used, the variant of the model using geometrical information was employed. In this paper, we also explore the differences, advantages and caveats of these two variants of the point-mass marginalization method.

\subsection{Mode projection: ``Project-out'' estimator}\label{sec:projout}

To further illustrate the similarities and/or differences among the above estimators, we also construct a new estimator that we refer to as the ``project-out'' estimator. This estimator is designed such that it follows the philosophy of the \citet{MacCrann2020} approach, namely by focusing on the point-mass $1/R^2$ mode within the observed galaxy-galaxy lensing data vector, while it follows the implementation used in \citet{Park2021}, namely by finding a suitable linear transformation and using it to transform the data vectors and covariances. In Appendix~\ref{sec:app_equivalence} we show that the "Proj-out" method is actually equivalent to the point-mass marginalization method when the prior of the point-mass is infinitely wide. Also, this basic equivalence between marginalizing over a parameter and projecting out a given mode had already been pointed out in Appendix A from \citet{Seljak1998} for a general scenario.

The core idea behind this estimator, thus, is to identify the projection of the observed galaxy-galaxy lensing data vector onto the $1/R^2$ mode, and then to remove it from the original vector. The projection operator $\mathbf{P}$ is given by \citep{aitken_1936, Tegmark1998}:
\begin{equation}
    \mathbf{P} = \mathbf{A} \left(\mathbf{A}^\mathbf{T} \mathbf{C}_\gamma^{-1} \mathbf{A}\right)^{-1} \mathbf{A}^\mathbf{T}\mathbf{C}_\gamma^{-1}.
\end{equation}
where $\mathbf{A}$ has columns spanning the subspace onto which we wish to project. In our case, $\mathbf{A}$ has only one column given by $\{1/R_i^2\}$, or in practice $\{1/\theta_i^2\}$, with $\theta_i$ being the representative angular separation of the $i$-th bin in our data vector. Note that instead of the ``vanilla'' projection operator $\mathbf{A} (\mathbf{A}^\mathbf{T} \mathbf{A})^{-1} \mathbf{A}^\mathbf{T}$ we use the generalized least squares definition of the projection operator to properly account for the covariances in the data vector.

The ``cleaned'' data vector is then defined as
\begin{eqnarray}
    \gamma_{t,\text{proj-out}} & = & \gamma_t - \mathbf{P}\gamma_t \nonumber \\
    & = & \left[\mathbf{I} - \mathbf{A} \left(\mathbf{A}^\mathbf{T}  \mathbf{C}_\gamma^{-1}  \mathbf{A}\right)^{-1} \mathbf{A}^\mathbf{T} \mathbf{C}_\gamma^{-1}\right] \gamma_t \nonumber \\
    & \equiv & \mathbf{M}\gamma_t,
\end{eqnarray}
and its covariance is given by 
\begin{equation}
    \mathbf{C}_\text{proj-out} = \mathbf{M} \mathbf{C}_\gamma \mathbf{M}^\mathbf{T}
\end{equation}
where $\mathbf{C}_\gamma$ is the original covariance matrix. The inversion of $\mathbf{C}_\text{proj-out}$ becomes problematic, as $\mathbf{M}$ is not a full-rank matrix. We thus follow \citet{Tegmark1997} to define a pseudoinverse of the transformed covariance matrix as 
\begin{equation}\label{eq:proj-out inverse cov}
    \tilde{\mathbf{C}}_\text{proj-out}^{-1} = \mathbf{M}\left(\mathbf{C}_\text{proj-out} + \eta \mathbf{A} \mathbf{A}^\mathbf{T} \right)^{-1} \mathbf{M}^\mathbf{T},
\end{equation}
where $\eta$ is a constant whose value does not affect the performance of the pseudoinverse\footnote{We have used $\eta=10^{-3}$ for the runs with DES Y3 data. We have checked that using a different value, e.g. $\eta=10^3$, does not change the results.}. This can be understood intuitively as adding back in the lost mode to $\mathbf{C}_\text{proj-out}$, inverting, and removing the added mode at the end. With $\gamma_{t,\text{proj-out}}$ and $\tilde{\mathbf{C}}_\text{proj-out}$ in hand, a full likelihood analysis can be defined using this new estimator.

\section{Methodology}\label{sec:methology}

The question we are aiming to answer is whether the different  methodologies to localize the galaxy-galaxy lensing measurements  are consistent under the precision of current and future surveys. In order to do so, we perform a combined 2$\times$2pt analysis using each of the methods to localize the tangential shear measurements and compare their performances at the cosmological posterior level. We choose to do the comparison in a 2$\times$2pt analysis instead of a whole 3$\times$2pt analysis to maximize the impact that the galaxy-galaxy lensing part of the data vector has on the cosmological parameter posteriors,  thus maximizing potential differences between the localizing estimators. We perform the comparison using two different setups: 1) First we assume the characteristics of a future survey to test the differences under the smallest covariance. In particular, we choose the specifications of a LSST Y1-like survey, since that will become relevant in the near-future, and it  will already be significantly more constraining than current generation surveys. 2) Secondly we  apply the comparison of the methods to  DES Y3 data, a noisy realistic scenario.

For the LSST Y1 simulated case, we generate the input data vectors using a \textit{contaminated} model that includes baryonic effects and non-linear galaxy bias contributions that mostly affect small scales (see Sec.~\ref{sec:contaminated-model}), and analyze it using the simpler \textit{fiducial} model that does not take into account these contributions with a linear galaxy bias model and a dark matter only power spectrum (see Sec.~\ref{sec:fiducialmodel}). We  find the appropriate set of scale cuts that allow us to recover unbiased cosmology in each case following the prescription described in Sec~\ref{sec:scale_cuts}.

\subsection{Fiducial model}
\label{sec:fiducialmodel}

Here we summarize the baseline or ``fiducial'' theory that we will use to model the observed tangential shear and galaxy clustering quantities. This is the same one used in the DES Y3 3$\times$2pt analysis. In particular, for the LSST Y1 simulated analysis described below we assume the model presented in more detail in \citet{y3-methods} and for the DES Y3 data analysis the one defined in \citet{y3-3x2pt}\footnote{The two models are in essence identical, but differ in some of the input parameters such as the  lens magnification parameters and the redshift distributions (which were measured later in the data analysis), and in  the priors of the nuisance parameters.}.

The tangential shear $\gamma_t$ and angular clustering $w(\theta)$ observables can be expressed as  various real space projections of  angular power spectra. In particular we model $\gamma_t$  as the following curved sky projection of the galaxy-matter angular power spectra and of other terms that encapsulate observational effects such as intrinsic alignments, lens magnification and their cross-talk, which add up to the total observed $C^{ij}_{\delta_{\mathrm{obs}} \, E}$:
\begin{equation}
\gamma_t^{ij}(\theta) = \sum_\ell \frac{2\ell+1}{4\pi\ell(\ell+1)} P^2_\ell(\cos\theta)  C^{ij}_{\delta_\mathrm{obs} \, E} ,
\end{equation}
and we model $w(\theta)$ as:
\begin{equation}
w^{ij}(\theta) = \sum_\ell \frac{2\ell+1}{4\pi}P_\ell(\cos\theta)   C^{ij}_{\delta_\mathrm{obs} \, \delta_\mathrm{obs}} . 
\end{equation}
We refer the reader to \citet{y3-methods} for a detailed definition of $C^{ij}_{\delta_{\mathrm{obs}}\,  E}$ and of  $C^{ij}_{\delta_\mathrm{obs}\, \delta_\mathrm{obs} }$.  $P_\ell$ and $P^2_\ell$ are the Legendre polynomials and the associated Legendre polynomials respectively. For the matter power spectrum we use the dark matter only Halofit prescription from \citet{Takahashi2012} and assume a linear galaxy bias to relate the galaxy and matter density fluctuations.

We obtain the LSST Y1 covariances with the publicly available \textsc{CosmoCov} code \citep{Krause2017, Fang2020}, using the number densities and noise specified in Tab.~\ref{tab:setup}. We include Gaussian and non-Gaussian terms computed using a halo model. We assume an area of 12300 deg$^2$ for LSST Y1, consistently with the specifications from the \citet{SRD}.

\begin{table}
\centering
\begin{tabular}{ccc}
\hline 
   & LSST Y1  \\
\hline 
\textbf{Lens Sample}  & \\
Number density  [arcmin$^{-2}$] & (2.29, 3.97, 6.06, 3.07, 2.62) \\
Galaxy bias &  (1.7, 1.7, 1.7, 2.0, 2.0)\\
\hline 

\textbf{Source Sample} &  \\
Number density [arcmin$^{-2}$] &    (2.50, 2.50, 2.52, 2.48) \\
Total shape noise & 0.3677 \\
\hline 
\bottomrule

\end{tabular}
\caption{LSST Y1 lens and source sample specifications in the setup of this work. The listed shape noise is the total one including both ellipticity components. These values are taken from \citet{SRD}, which specifies a lens number density of 18 arcmin$^{-2}$, which we split in 5 redshift bins, and a source number density of 10 arcmin$^{-2}$  which we split in 4 redshift bins. }\label{tab:setup}
\end{table}

We use a $\Lambda$CDM model with 5 (6) free cosmological  parameters for the simulated (data) case: $\Omega_m$, $A_s$, $n_s$, $\Omega_b$, $h$, ($\Omega_\nu$). We also free  additional nuisance parameters to marginalize over uncertainties related to photometric redshifts -- both for the lens and source samples,  intrinsic alignments and shear calibration. The full list of free parameters and their respective priors can be found in table~II\footnote{The only difference between the values that we use and the ones from table II from \citet{y3-methods} is that we fix the neutrino density parameter $\Omega_\nu h^2$ to a null value. This is because \textsc{CosmoCov} is not able to generate a model for the covariance that takes into account neutrinos. However, as shown in figure 2 from  \citet{y3-methods}, the impact of marginalizing over neutrino density is small for the DES Y3 3$\times$2pt analysis, so we do not expect this choice to affect any of the conclusions of this work.} from \citet{y3-methods} for the simulated analysis and in table~I from \citet{y3-3x2pt} for the DES Y3 data analysis. For the simulated LSST Y1 analysis we  assume the same redshift distributions that were used in the methodology paper of the DES Y3 analysis \citep{y3-methods}, which are an early estimate of the DES Y3 redshift distributions\footnote{LSST is expected to use a different redshift binning with respect to the one we choose in this work. However, since we are matching the rest of the settings to \citet{y3-methods} we decided to also match the input redshift distributions for simplicity, also given the fact that the redshift distributions that we assume in this work have a comparable binning  and redshift range than the one predicted in \citet{SRD} for the LSST Y1 sample.}. For the DES Y3 data chains, we use the same settings and priors as in \citet{y3-3x2pt}, except that we do not include the shear-ratio likelihood and only combine the tangential shear and galaxy clustering measurements, since for the 2$\times$2pt case it does not significantly change the results \citep*{desy3-shearratio}. 

We use the \textsc{CosmoSIS} \citep{Zuntz2015} framework to generate the data vectors and perform the 2$\times$2pt analysis. We use \textsc{MultiNest} \citep{Feroz2008, Feroz2009, Feroz2019}  to sample the parameter space and obtain the parameter posteriors, with the following accuracy settings: \texttt{live\_points=500}, \texttt{efficiency=0.3}, \texttt{tolerance=0.01}, \texttt{constant\_efficiency=F}. For the DES Y3 data,  we use \textsc{Polychord}  with the same high-accuracy settings  used in \citet{y3-3x2pt}. 

\subsection{Contaminated input model} \label{sec:contaminated-model}

The baryonic contamination is obtained by rescaling the non-linear matter power spectrum  with the baryonic effects from OWLS (OverWhelmingly Large Simulations project, \citealt{OWLS, vanDaalen11}) as a function of redshift and scale, comparing the power spectrum from the dark matter-only simulation with the power spectrum from the OWLS AGN simulation, following \citet{y3-methods}.   For the non-linear galaxy bias contamination we utilize a model that has been calibrated using $N$-body simulations and is described in \citet{Pandey_2020,y3-2x2ptbias}. Overall we use the same procedure which is used in \citet{y3-methods} to contaminate the fiducial data vector with these effects. Note that while the scale cuts and constraining power for each setup will depend on these choices (both on the contaminated and fiducial model), the comparison of the estimators will be independent of it since we use the same input contamination for all the different localization methodologies.

\begin{figure}
\begin{center}
\includegraphics[width=0.49\textwidth]{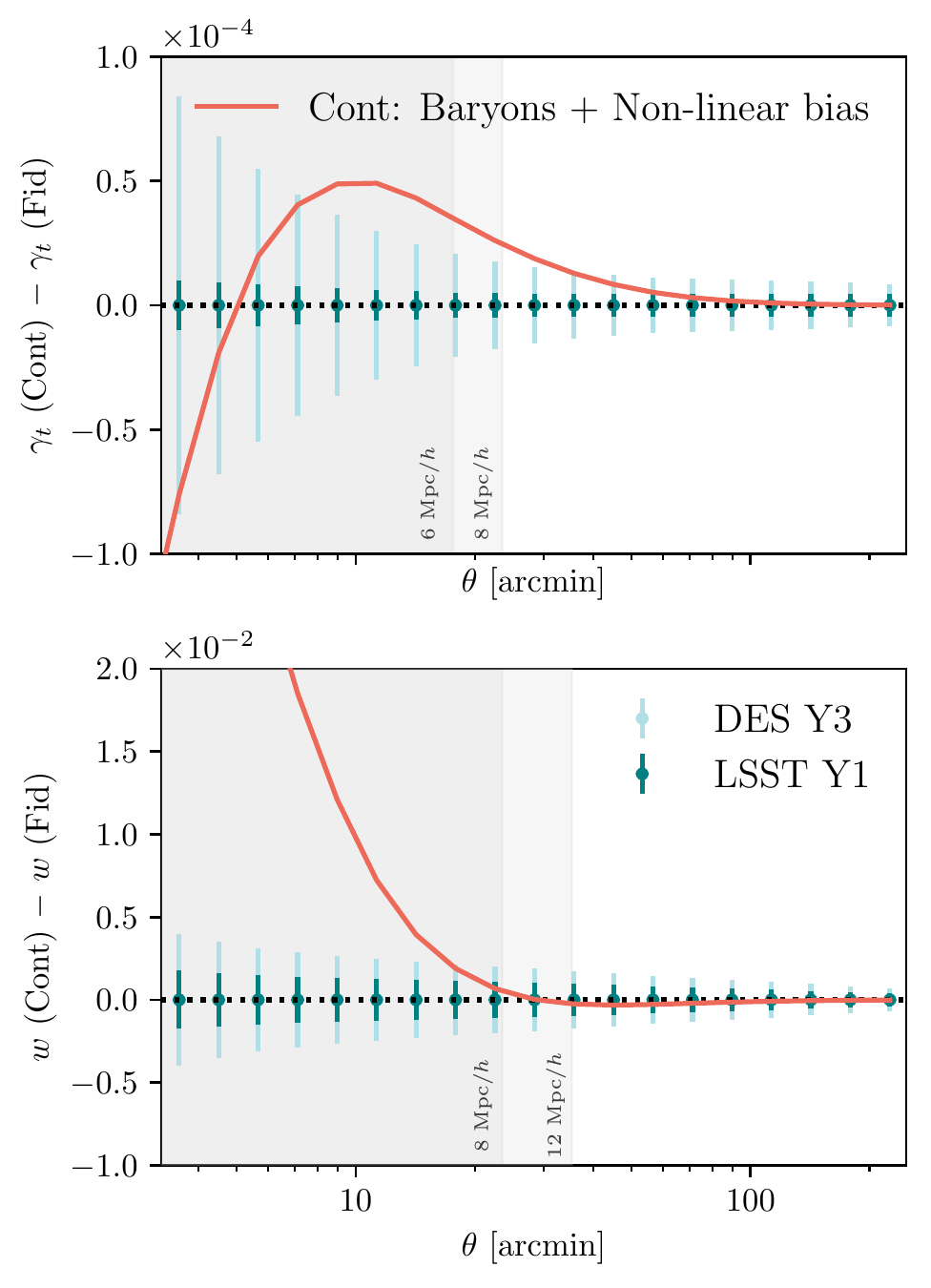}
\end{center}
\caption{Difference between the contaminated data vector and the fiducial ones for the tangential shear (top) and angular clustering (bottom), for the redshift bin combination corresponding to the second lens bin and the highest source bin. This redshift bin combination corresponds to one of the higher S/N ones.  Green error bars represent the uncertainties for DES Y3 and LSST Y1. The gray regions mark the scale cuts that are needed to obtain  unbiased cosmological results from this contamination,  which have been determined to be $w>8$ Mpc/$h$ and $\gamma_t>6$ Mpc/$h$ for DES Y3 and  $w>12$ Mpc/$h$ and $\gamma_t>8$ Mpc/$h$ for LSST Y1 when using one of the methods to localize the tangential shear measurements (otherwise the scale cuts would need to be larger for the tangential shear quantity, as shown in Appendix~\ref{app:no-mitigation})}. 
\label{fig:contaminated_datavecs}
\end{figure}

We generate both the fiducial and contaminated data vectors at the same cosmological and nuisance parameters that were used to define the scale cuts in the DES Y3 3$\times$2pt cosmological analysis. In Fig.~\ref{fig:contaminated_datavecs} we display the differences between the contaminated and fiducial data vectors for the tangential shear in the top and angular galaxy clustering in the bottom.

\subsection{Procedure to obtain the scale cuts}\label{sec:scale_cuts}

Here we describe how we obtain the scale cuts that we can use for LSST Y1 that yield unbiased cosmological results given our input contamination data vectors. We compute the differences of the posteriors in the 2D $S_8$--$\Omega_m$ plane between results using either the fiducial or the contaminated input data vectors. Specifically, we use the maximum a posteriori point (MAP) to compute the 2D offsets. We choose to use the "Point-mass (free per $z_l^i$)" model to perform this exercise, although we do not expect this choice to impact the results for the scale cuts. If the difference is above 0.3$\sigma$, it does not pass our criteria, following the same procedure as in \citet{y3-methods}. We have tested the following set of scale cuts: $w>8$ Mpc/$h$, $\gamma_t > 8$ Mpc/$h$; $w>12$ Mpc/$h$, $\gamma_t > 6$ Mpc/$h$;  $w>12$ Mpc/$h$, $\gamma_t > 8$ Mpc/$h$. Only this last set of scale cuts meets the criteria. See Appendix~\ref{app:scale-cuts} for the plots showing these differences.

\section{Results}
\label{sec:results}

\begin{figure*}
\begin{center}
\includegraphics[width=0.7\textwidth]{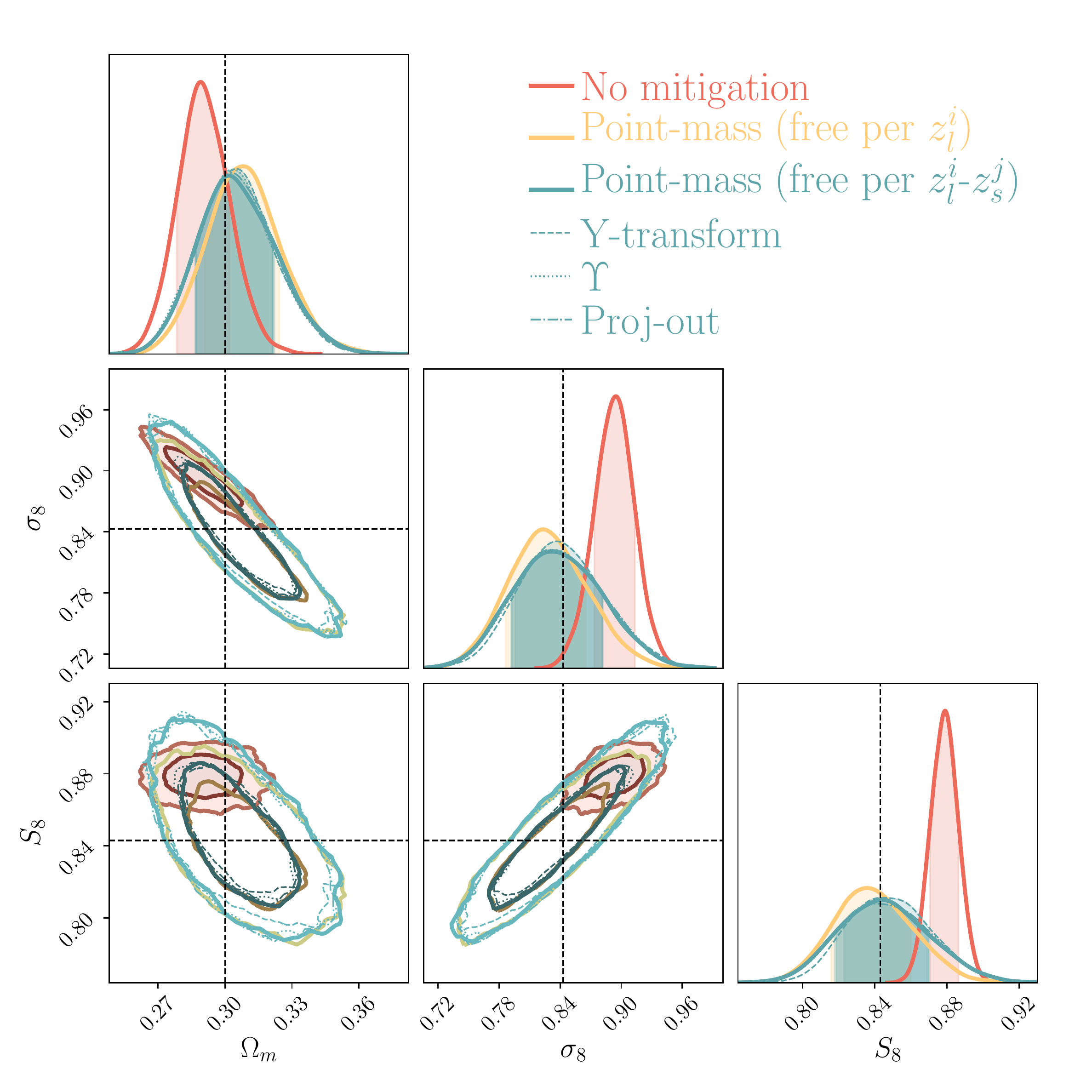}
\caption{Cosmological parameter posteriors obtained from an input galaxy clustering and  galaxy-galaxy lensing data vector (2$\times$2pt)  with non-linear bias and baryonic contamination, LSST Y1 covariance, 8Mpc/$h$ scale cuts for galaxy-galaxy lensing and 12 Mpc/$h$ for galaxy clustering. This figure demonstrates that all the methodologies to localize the galaxy-galaxy lensing measurements  perform similarly at the cosmological posterior level with LSST Y1 uncertainties.
The 2D contours represent 1$\sigma$ and 2$\sigma$ confidence regions. The shaded area under the 1D posteriors represents the 1$\sigma$ confidence level in 1D.} 
\label{fig:cosmo_estimators_comparison}
\end{center}
\end{figure*}

Using the LSST Y1 setup described in the previous section, we find that all the estimators perform in a similar way. In Fig.~\ref{fig:cosmo_estimators_comparison} we show the results for the simulated $2\times 2$pt analysis, combining galaxy-galaxy lensing and galaxy clustering for all the methodologies that we want to compare that localize the tangential shear measurements. We also add the result without applying any mitigation method, to illustrate the importance of using one of these methodologies to obtain unbiased cosmological constraints. All these results are applying  the fiducial scale cuts that passed the criteria defined in Sec.~\ref{sec:scale_cuts}: $w>12$ Mpc/$h$ and $\gamma_t > 8$ Mpc/$h$. In Appendix~\ref{app:no-mitigation} we also show that $\gamma_t > 40$ Mpc/$h$ cuts  would be needed in order to recover unbiased cosmological constraints if we do not apply any mitigation scheme. 

We find that all the methodologies are able to properly mitigate the impact of the input contamination and recover very similar uncertainties on the most constrained cosmological parameters of a 2$\times$2pt analysis, that is, $\Omega_m$ and $\sigma_8$. In the comparison we also include the new estimator that we have developed that \textit{only} projects out the $1/R^2$ mode, without doing any extra transformation as in some of the other methodologies, labelled as "Proj-out" in the plot. The fact that all the estimators agree with each other, and also with this new estimator, indicates that projecting out this mode is the only thing that has any effect in all the mitigation methodologies at the cosmological posterior level. 

Instead we have found that  differences between the methods arise from input assumptions. In particular, we observe the biggest difference is between the two different variants of the point-mass marginalization. The method labelled as "Point-mass (free per $z^i_l-z_s^j$)" does not use any extra information with respect to the other estimators and can be compared directly to them. On the other hand the one labelled as "Point-mass (free per $z^i_l$)" uses the approximation that the mass enclosed below the minimum scale used does not evolve within the redshift range of the lens bin, and moreover uses geometrical "shear-ratio" information to constrain the scaling between different sources sharing the same lens bin.  We find that the posteriors for this case are slightly more constraining as expected since they use more information, but also slightly more biased with respect to the input true cosmology. Thus, we recommend that when applying the point-mass marginalization case using geometrical information  to LSST Y1 or a more constraining data set, the assumption that the point mass evolves slowly within a lens redshift bin should be tested for a given lens sample.

\begin{figure*}
\begin{center}
\includegraphics[width=0.7\textwidth]{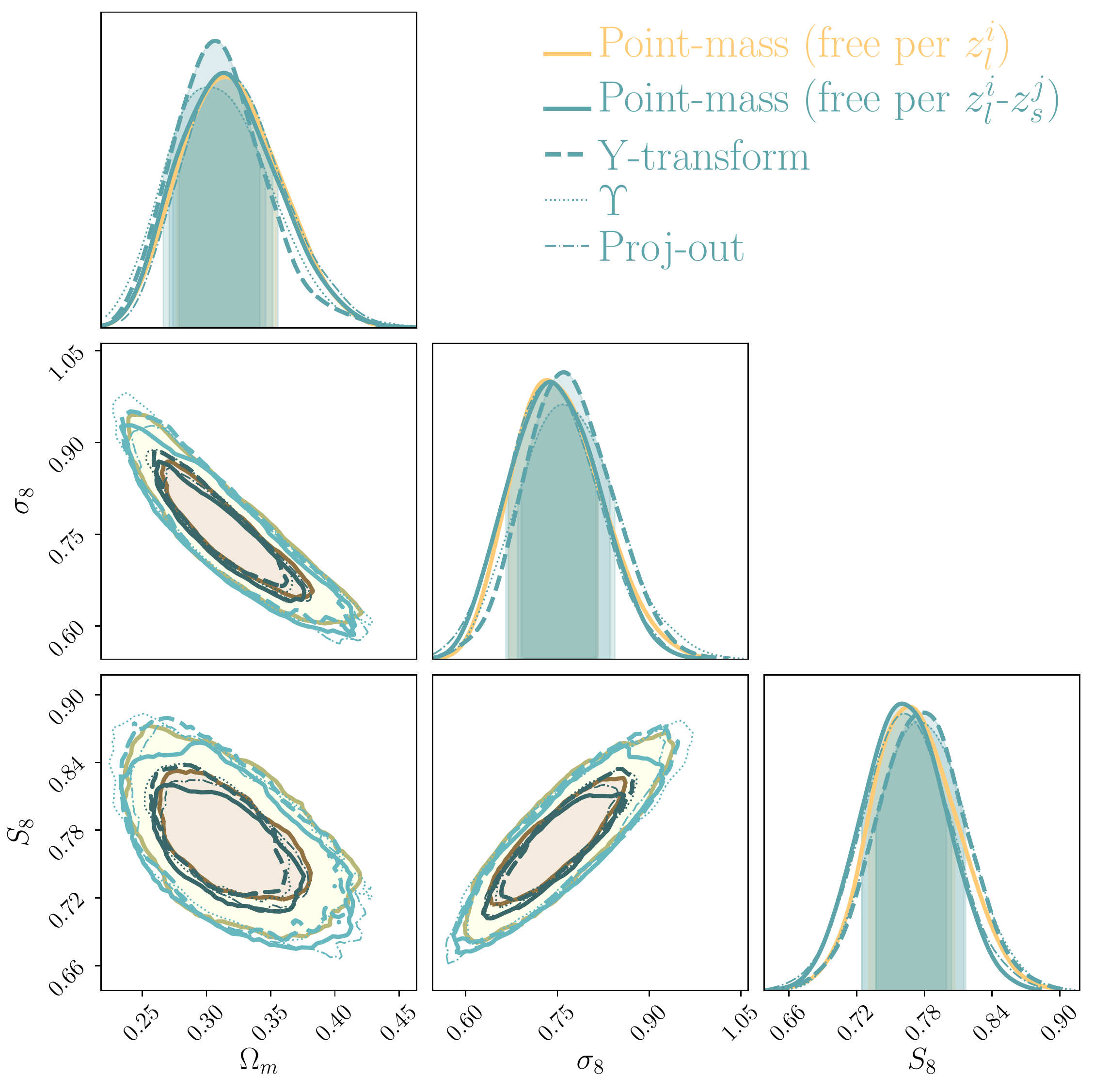}
\end{center}
\caption{Application to DES Y3 data in a 2$\times$2pt analysis for the \textsc{MagLim} sample of each of the different methodologies to localize the tangential shear measurements. This figure demonstrates that all the methodologies to localize the galaxy-galaxy lensing measurements  perform similarly at the cosmological posterior level with  DES Y3 uncertainties and the presence of noise. The 2D contours represent 1$\sigma$ and 2$\sigma$ confidence regions. The shaded area under the 1D posteriors  represent the 1$\sigma$ confidence level in 1D.} 
\label{fig:desy3}
\end{figure*}

In Appendix~\ref{app:IA} we also compare the intrinsic alignment parameter posteriors. We find that all the estimators perform similarly, except for the point-mass using geometric information to constrain the scaling between the lens redshift bins that gives a much tighter (and still unbiased) posterior on the galaxy bias of the source sample $b_{TA}$ parameter and a slightly tighter constraint for the $A_2$ and $\alpha_2$ parameters, which are the parameters related to the tidal torque contribution to the model.

\subsection{Application to DES Y3 data}

After testing on noiseless simulated data vector with an LSST Y1 setup we apply all the methodologies to localize the galaxy-galaxy lensing measurements to the DES Y3 data, in particular to the  2$\times$2pt setup with the \textsc{MagLim} lens sample presented in \cite{desy3-2x2pt-maglim}
but  without the shear-ratio likelihood. The scale cuts we use are the same as in \cite{desy3-2x2pt-maglim}, that is, 6 Mpc/$h$ for galaxy-galaxy lensing and 8 Mpc/$h$ for the angular galaxy clustering. In Fig.~\ref{fig:desy3} we show the results of this comparison. Note that the "Point-mass (free per $z^i_l$)" case corresponds to the fiducial DES Y3 2$\times$2pt result presented in \cite{desy3-2x2pt-maglim}. We find that all the methodologies give consistent results, even in this noisy and more realistic scenario, which presents non-linearities at the small scales and includes all the effects from the real Universe. In Appendix~\ref{app:desy3vslssty1} we compare the constraining power of DES Y3 and LSST Y1 to give a sense of scale.

Moreover, we  compare the posteriors on the TATT intrinsic alignment parameters for DES Y3 data and find similar conclusions as in the simulated case, as shown in Appendix~\ref{app:IA}.

\subsection{Performance differences}\label{sec:performance}

Regarding performance differences between the methods, we find the "Proj-out" estimator to be less numerically stable than the other approaches in its current \textsc{CosmoSIS} implementation. This is because the "Proj-out" method requires an input arbitrary $\eta$ value to obtain the pseudoinverse of the transformed covariance matrix, as defined in Eq.~(\ref{eq:proj-out inverse cov}). While we have checked that a considerable wide range of $\eta$  values yield the same results under the conditions of this analysis, outside a certain range that is no longer the case. Thus, robustness against different $\eta$ values might need to be revisited in other settings. Moreover, we find the point-mass marginalization method to be the simplest to use as currently implemented in  \textsc{CosmoSIS}, since it does not require modifying the input data vector or covariance matrix files. Finally, we also compare the running time of the different estimators. Using the LSST Y1 setup, the point-mass marginalization without geometrical information took 7h, the $\Upsilon$ statistic took 7h 45 min, the Y-transformation took 12h and the proj-out estimator took 16h 40 min, using the \textsc{MultiNest} sampler with the settings defined above and using the same number of cores. On DES Y3 data and the \textsc{Polychord} sampler with high-accuracy settings, the point-mass marginalization without geometrical information took 45h, the $\Upsilon$ statistic took 49h, the Y-transformation took 54h and the proj-out estimator took 70h. We summarize these findings in Table~\ref{tab:comparison}.

\begin{table*}
\centering
\begin{tabular}{ccccc}
\hline 
  & Modify data vector? & Numerical Stability & Computational expense  \\
\hline 
Point-mass marginalization &  No &   Excellent & Fastest  \\
Annular differential surface density ($\Upsilon$) & Yes & Excellent &  Fast\\
Y-transformation &  Yes & Excellent & Slow \\
Proj-out &  Yes  & Poor & Slowest \\
\bottomrule
\end{tabular}
\caption{Comparison of the process and performance of each methodology to localize the galaxy-galaxy lensing measurements. The computational expense estimates are based on the current implementation of the \textsc{CosmoSIS} code. More details about the performance differences can be found in Sec.~\ref{sec:performance}.
}\label{tab:comparison}
\end{table*}

\section{Summary and conclusions}
\label{sec:summary}

In this paper we compare three existing  methodologies to localize the galaxy-galaxy lensing measurements: the Annular Differential Surface Density estimator ($\Upsilon$) presented in \citet{Baldauf2010}, the Y-transformation  derived in \citet{Park2021} and the point-mass marginalization described in \citet{MacCrann2020}. We compare them at the cosmological posterior level, performing a 2$\times$2pt analysis which combines projected angular clustering and tangential shear measurements. We find that all these methods  are able to mitigate the impact of small scale information when using a LSST Y1  setup with noiseless simulated  data vectors, and that they are all performing in a very similar manner, yielding equivalent posteriors on the cosmological parameters. 

To further illustrate the similarities and/or differences  amongst the above listed estimators, we also construct a new estimator that we refer to as "project-out". The "project-out" method identifies the projection of the observed galaxy-galaxy lensing data vector onto the $1/R^2$ mode, and then it removes it from the original vector, following a similar procedure to the Y-transform methodology, but in this case \textit{only} removing this mode. Then, we proceed to compare the posteriors obtained with the "project-out" method to the other ones, finding it yields equivalent results. Therefore, we conclude that the removal of the $1/R^2$ mode is the only relevant transformation that is needed to localize the tangential shear measurements and that the rest of the modifications in the other estimators are not producing any significant differences at the cosmological posterior level. 

We also compare two different variations of the point-mass marginalization methodology, one that uses exactly the same information as the other estimators and one that uses extra geometrical information to constrain the scaling of the point-mass between different lens and source bin combinations, by assuming that the enclosed mass does not evolve with redshift within the width of the lens bin. We find that the point-mass marginalization using geometric information yields slightly more constraining but also slightly  biased results on the cosmological parameters in the LSST Y1 simulated case, due to the approximation it makes. Thus, the  assumption going into this point-mass variant should always be tested before applying it to  more constraining data sets.
Notably, we also find that the extra geometrical information significantly improves the precision (while keeping the  accuracy) of the intrinsic alignment parameters of the tidal alignment and tidal torque (TATT) model. In particular, we find the biggest difference in the posterior for the galaxy bias of the source sample $b_{\text{TA}}$  and in the parameters controlling the tidal torque part of the TATT model. 

We also compare the results obtained using any of the mitigation schemes with the case of not applying any mitigation scheme but applying larger scale cuts. With the LSST Y1 setup, we find that the mitigation schemes yield $\sim$1.3 times more constraining $S_8$ results than applying  larger scale cuts without using any mitigation scheme.

Finally, we apply all the methods to DES Y3 data, reaching very similar conclusions as in the simulated case. However, even if the DES Y3 data has larger uncertainties than the simulated LSST Y1 case, this exercise is still meaningful since it provides an input data vector with the non-linearities and baryonic effects of the real Universe, together with any other other unforeseen contamination that is not present in our fiducial model. It also tests the methods in the presence of noise. In this case we still find that all the methodologies perform in a similar manner.

\section*{Acknowledgments}
Funding for the DES Projects has been provided by the U.S. Department of Energy, the U.S. National Science Foundation, the Ministry of Science and Education of Spain, 
the Science and Technology Facilities Council of the United Kingdom, the Higher Education Funding Council for England, the National Center for Supercomputing 
Applications at the University of Illinois at Urbana-Champaign, the Kavli Institute of Cosmological Physics at the University of Chicago, 
the Center for Cosmology and Astro-Particle Physics at the Ohio State University,
the Mitchell Institute for Fundamental Physics and Astronomy at Texas A\&M University, Financiadora de Estudos e Projetos, 
Funda{\c c}{\~a}o Carlos Chagas Filho de Amparo {\`a} Pesquisa do Estado do Rio de Janeiro, Conselho Nacional de Desenvolvimento Cient{\'i}fico e Tecnol{\'o}gico and 
the Minist{\'e}rio da Ci{\^e}ncia, Tecnologia e Inova{\c c}{\~a}o, the Deutsche Forschungsgemeinschaft and the Collaborating Institutions in the Dark Energy Survey. 

The Collaborating Institutions are Argonne National Laboratory, the University of California at Santa Cruz, the University of Cambridge, Centro de Investigaciones Energ{\'e}ticas, 
Medioambientales y Tecnol{\'o}gicas-Madrid, the University of Chicago, University College London, the DES-Brazil Consortium, the University of Edinburgh, 
the Eidgen{\"o}ssische Technische Hochschule (ETH) Z{\"u}rich, 
Fermi National Accelerator Laboratory, the University of Illinois at Urbana-Champaign, the Institut de Ci{\`e}ncies de l'Espai (IEEC/CSIC), 
the Institut de F{\'i}sica d'Altes Energies, Lawrence Berkeley National Laboratory, the Ludwig-Maximilians Universit{\"a}t M{\"u}nchen and the associated Excellence Cluster Universe, 
the University of Michigan, NSF's NOIRLab, the University of Nottingham, The Ohio State University, the University of Pennsylvania, the University of Portsmouth, 
SLAC National Accelerator Laboratory, Stanford University, the University of Sussex, Texas A\&M University, and the OzDES Membership Consortium.

Based in part on observations at Cerro Tololo Inter-American Observatory at NSF's NOIRLab (NOIRLab Prop. ID 2012B-0001; PI: J. Frieman), which is managed by the Association of Universities for Research in Astronomy (AURA) under a cooperative agreement with the National Science Foundation.

The DES data management system is supported by the National Science Foundation under Grant Numbers AST-1138766 and AST-1536171.
The DES participants from Spanish institutions are partially supported by MICINN under grants ESP2017-89838, PGC2018-094773, PGC2018-102021, SEV-2016-0588, SEV-2016-0597, and MDM-2015-0509, some of which include ERDF funds from the European Union. IFAE is partially funded by the CERCA program of the Generalitat de Catalunya.
Research leading to these results has received funding from the European Research
Council under the European Union's Seventh Framework Program (FP7/2007-2013) including ERC grant agreements 240672, 291329, and 306478.
We  acknowledge support from the Brazilian Instituto Nacional de Ci\^encia
e Tecnologia (INCT) do e-Universo (CNPq grant 465376/2014-2).

This manuscript has been authored by Fermi Research Alliance, LLC under Contract No. DE-AC02-07CH11359 with the U.S. Department of Energy, Office of Science, Office of High Energy Physics.

\section*{Data availability}

The data underlying this article are available in the DES Y3 Cosmology Data Release page at \texttt{https://des.ncsa.illinois.edu/releases/y3a2/Y3key-products}. The code developed for this project is integrated within the \textsc{CosmoSIS} framework and can be shared upon request.

\bibliography{library}
\bibliographystyle{mnras_2author}

\section*{Affiliations}
$^{1}$ Department of Astronomy and Astrophysics, University of Chicago, Chicago, IL 60637, USA\\
$^{2}$ Kavli Institute for Cosmological Physics, University of Chicago, Chicago, IL 60637, USA\\
$^{3}$ Kavli Institute for the Physics and Mathematics of the Universe (WPI), UTIAS, The University of Tokyo, Kashiwa, Chiba 277-8583, Japan\\
$^{4}$ Department of Applied Mathematics and Theoretical Physics, University of Cambridge, Cambridge CB3 0WA, UK\\
$^{5}$ NASA Goddard Space Flight Center, Greenbelt, MD 20771, USA\\
$^{6}$ Department of Physics and Astronomy, University of Pennsylvania, Philadelphia, PA 19104, USA\\
$^{7}$ Department of Physics, Northeastern University, Boston, MA 02115, USA\\
$^{8}$ Instituci\'o Catalana de Recerca i Estudis Avan\c{c}ats, E-08010 Barcelona, Spain\\
$^{9}$ Institut de F\'{\i}sica d'Altes Energies (IFAE), The Barcelona Institute of Science and Technology, Campus UAB, 08193 Bellaterra (Barcelona) Spain\\
$^{10}$ Argonne National Laboratory, 9700 South Cass Avenue, Lemont, IL 60439, USA\\
$^{11}$ Department of Physics, University of Michigan, Ann Arbor, MI 48109, USA\\
$^{12}$ Institute of Astronomy, University of Cambridge, Madingley Road, Cambridge CB3 0HA, UK\\
$^{13}$ Kavli Institute for Cosmology, University of Cambridge, Madingley Road, Cambridge CB3 0HA, UK\\
$^{14}$ Physics Department, 2320 Chamberlin Hall, University of Wisconsin-Madison, 1150 University Avenue Madison, WI  53706-1390\\
$^{15}$ Department of Physics, Duke University Durham, NC 27708, USA\\
$^{16}$ California Institute of Technology, 1200 East California Blvd, MC 249-17, Pasadena, CA 91125, USA\\
$^{17}$ Instituto de F\'{i}sica Te\'orica, Universidade Estadual Paulista, S\~ao Paulo, Brazil\\
$^{18}$ Laborat\'orio Interinstitucional de e-Astronomia - LIneA, Rua Gal. Jos\'e Cristino 77, Rio de Janeiro, RJ - 20921-400, Brazil\\
$^{19}$ Department of Physics, Carnegie Mellon University, Pittsburgh, Pennsylvania 15312, USA\\
$^{20}$ Instituto de Astrofisica de Canarias, E-38205 La Laguna, Tenerife, Spain\\
$^{21}$ Universidad de La Laguna, Dpto. Astrofísica, E-38206 La Laguna, Tenerife, Spain\\
$^{22}$ Center for Astrophysical Surveys, National Center for Supercomputing Applications, 1205 West Clark St., Urbana, IL 61801, USA\\
$^{23}$ Department of Astronomy, University of Illinois at Urbana-Champaign, 1002 W. Green Street, Urbana, IL 61801, USA\\
$^{24}$ Physics Department, William Jewell College, Liberty, MO, 64068\\
$^{25}$ Jodrell Bank Center for Astrophysics, School of Physics and Astronomy, University of Manchester, Oxford Road, Manchester, M13 9PL, UK\\
$^{26}$ Institut d'Estudis Espacials de Catalunya (IEEC), 08034 Barcelona, Spain\\
$^{27}$ Institute of Space Sciences (ICE, CSIC),  Campus UAB, Carrer de Can Magrans, s/n,  08193 Barcelona, Spain\\
$^{28}$ Kavli Institute for Particle Astrophysics \& Cosmology, P. O. Box 2450, Stanford University, Stanford, CA 94305, USA\\
$^{29}$ Lawrence Berkeley National Laboratory, 1 Cyclotron Road, Berkeley, CA 94720, USA\\
$^{30}$ Fermi National Accelerator Laboratory, P. O. Box 500, Batavia, IL 60510, USA\\
$^{31}$ NSF AI Planning Institute for Physics of the Future, Carnegie Mellon University, Pittsburgh, PA 15213, USA\\
$^{32}$ Universit\'e Grenoble Alpes, CNRS, LPSC-IN2P3, 38000 Grenoble, France\\
$^{33}$ Department of Astronomy/Steward Observatory, University of Arizona, 933 North Cherry Avenue, Tucson, AZ 85721-0065, USA\\
$^{34}$ Jet Propulsion Laboratory, California Institute of Technology, 4800 Oak Grove Dr., Pasadena, CA 91109, USA\\
$^{35}$ Center for Cosmology and Astro-Particle Physics, The Ohio State University, Columbus, OH 43210, USA\\
$^{36}$ Department of Physics, The Ohio State University, Columbus, OH 43210, USA\\
$^{37}$ Department of Astronomy, University of California, Berkeley,  501 Campbell Hall, Berkeley, CA 94720, USA\\
$^{38}$ University Observatory, Faculty of Physics, Ludwig-Maximilians-Universit\"at, Scheinerstr. 1, 81679 Munich, Germany\\
$^{39}$ School of Physics and Astronomy, Cardiff University, CF24 3AA, UK\\
$^{40}$ Department of Astronomy, University of Geneva, ch. d'\'Ecogia 16, CH-1290 Versoix, Switzerland\\
$^{41}$ Department of Physics, University of Arizona, Tucson, AZ 85721, USA\\
$^{42}$ Department of Physics, Stanford University, 382 Via Pueblo Mall, Stanford, CA 94305, USA\\
$^{43}$ SLAC National Accelerator Laboratory, Menlo Park, CA 94025, USA\\
$^{44}$ Instituto de F\'isica Gleb Wataghin, Universidade Estadual de Campinas, 13083-859, Campinas, SP, Brazil\\
$^{45}$ Institute for Astronomy, University of Edinburgh, Edinburgh EH9 3HJ, UK\\
$^{46}$ ICTP South American Institute for Fundamental Research\\ Instituto de F\'{\i}sica Te\'orica, Universidade Estadual Paulista, S\~ao Paulo, Brazil\\
$^{47}$ Centro de Investigaciones Energ\'eticas, Medioambientales y Tecnol\'ogicas (CIEMAT), Madrid, Spain\\
$^{48}$ Brookhaven National Laboratory, Bldg 510, Upton, NY 11973, USA\\
$^{49}$ Department of Physics and Astronomy, Stony Brook University, Stony Brook, NY 11794, USA\\
$^{50}$ D\'{e}partement de Physique Th\'{e}orique and Center for Astroparticle Physics, Universit\'{e} de Gen\`{e}ve, 24 quai Ernest Ansermet, CH-1211 Geneva, Switzerland\\
$^{51}$ Excellence Cluster Origins, Boltzmannstr.\ 2, 85748 Garching, Germany\\
$^{52}$ Max Planck Institute for Extraterrestrial Physics, Giessenbachstrasse, 85748 Garching, Germany\\
$^{53}$ Universit\"ats-Sternwarte, Fakult\"at f\"ur Physik, Ludwig-Maximilians Universit\"at M\"unchen, Scheinerstr. 1, 81679 M\"unchen, Germany\\
$^{54}$ Cerro Tololo Inter-American Observatory, NSF's National Optical-Infrared Astronomy Research Laboratory, Casilla 603, La Serena, Chile\\
$^{55}$ Institute of Cosmology and Gravitation, University of Portsmouth, Portsmouth, PO1 3FX, UK\\
$^{56}$ CNRS, UMR 7095, Institut d'Astrophysique de Paris, F-75014, Paris, France\\
$^{57}$ Sorbonne Universit\'es, UPMC Univ Paris 06, UMR 7095, Institut d'Astrophysique de Paris, F-75014, Paris, France\\
$^{58}$ Department of Physics \& Astronomy, University College London, Gower Street, London, WC1E 6BT, UK\\
$^{59}$ Astronomy Unit, Department of Physics, University of Trieste, via Tiepolo 11, I-34131 Trieste, Italy\\
$^{60}$ INAF-Osservatorio Astronomico di Trieste, via G. B. Tiepolo 11, I-34143 Trieste, Italy\\
$^{61}$ Institute for Fundamental Physics of the Universe, Via Beirut 2, 34014 Trieste, Italy\\
$^{62}$ Hamburger Sternwarte, Universit\"{a}t Hamburg, Gojenbergsweg 112, 21029 Hamburg, Germany\\
$^{63}$ Department of Physics, IIT Hyderabad, Kandi, Telangana 502285, India\\
$^{64}$ Institute of Theoretical Astrophysics, University of Oslo. P.O. Box 1029 Blindern, NO-0315 Oslo, Norway\\
$^{65}$ Department of Astronomy, University of Michigan, Ann Arbor, MI 48109, USA\\
$^{66}$ School of Mathematics and Physics, University of Queensland,  Brisbane, QLD 4072, Australia\\
$^{67}$ Santa Cruz Institute for Particle Physics, Santa Cruz, CA 95064, USA\\
$^{68}$ Center for Astrophysics $\vert$ Harvard \& Smithsonian, 60 Garden Street, Cambridge, MA 02138, USA\\
$^{69}$ Departamento de F\'isica Matem\'atica, Instituto de F\'isica, Universidade de S\~ao Paulo, CP 66318, S\~ao Paulo, SP, 05314-970, Brazil\\
$^{70}$ Observat\'orio Nacional, Rua Gal. Jos\'e Cristino 77, Rio de Janeiro, RJ - 20921-400, Brazil\\
$^{71}$ Department of Astrophysical Sciences, Princeton University, Peyton Hall, Princeton, NJ 08544, USA\\
$^{72}$ \\
$^{73}$ School of Physics and Astronomy, University of Southampton,  Southampton, SO17 1BJ, UK\\
$^{74}$ Computer Science and Mathematics Division, Oak Ridge National Laboratory, Oak Ridge, TN 37831\\

\appendix

\section{Equivalence between "Project-out" and Point-mass marginalization} \label{sec:app_equivalence}

In this appendix we show that the "Project-out"  and the Point-mass marginalization methods are mathematically equivalent if the prior for the point-mass is infinitely wide. Let us define $\mathbf{m}$ to be the model prediction and $\mathbf{d}$ to be the data vector. 

For the PM marginalization case we have:
$$
\chi^2_{\rm PM} = (\mathbf{m}-\mathbf{d})^T \mathbf{N}^{-1} (\mathbf{m}-\mathbf{d}) \; .
$$
However, in the infinite-prior case it is:
$$
\mathbf{N}^{-1} = \mathbf{C}_\gamma^{-1} - \mathbf{C}_\gamma^{-1} \mathbf{A} (\mathbf{A}^T\mathbf{C}_\gamma^{-1}\mathbf{A})^{-1} \mathbf{A}^T \mathbf{C}_\gamma^{-1} = \mathbf{C}_\gamma^{-1} (\mathbf{I}-\mathbf{P}) \; ,
$$
which results in 
\begin{align}\label{eq:PMchi2}
    \chi^2_{\rm PM} = (\mathbf{m}-\mathbf{d})^T \mathbf{C}_\gamma^{-1} \mathbf{M} (\mathbf{m}-\mathbf{d}) \; .
\end{align}
So, $\mathbf{N}^{-1}(\mathbf{m}-\mathbf{d})$ exactly simply removes the contribution from the $1/\theta^2$ mode from the model and data vectors, like the project-out approach. The project-out method is by construction only removing that mode from the data vector and does nothing else. Therefore, the two methods are equivalent in the above limit. 

\section{Scale cuts}\label{app:scale-cuts}

\begin{figure}
\centering
 \includegraphics[width=0.49\textwidth]{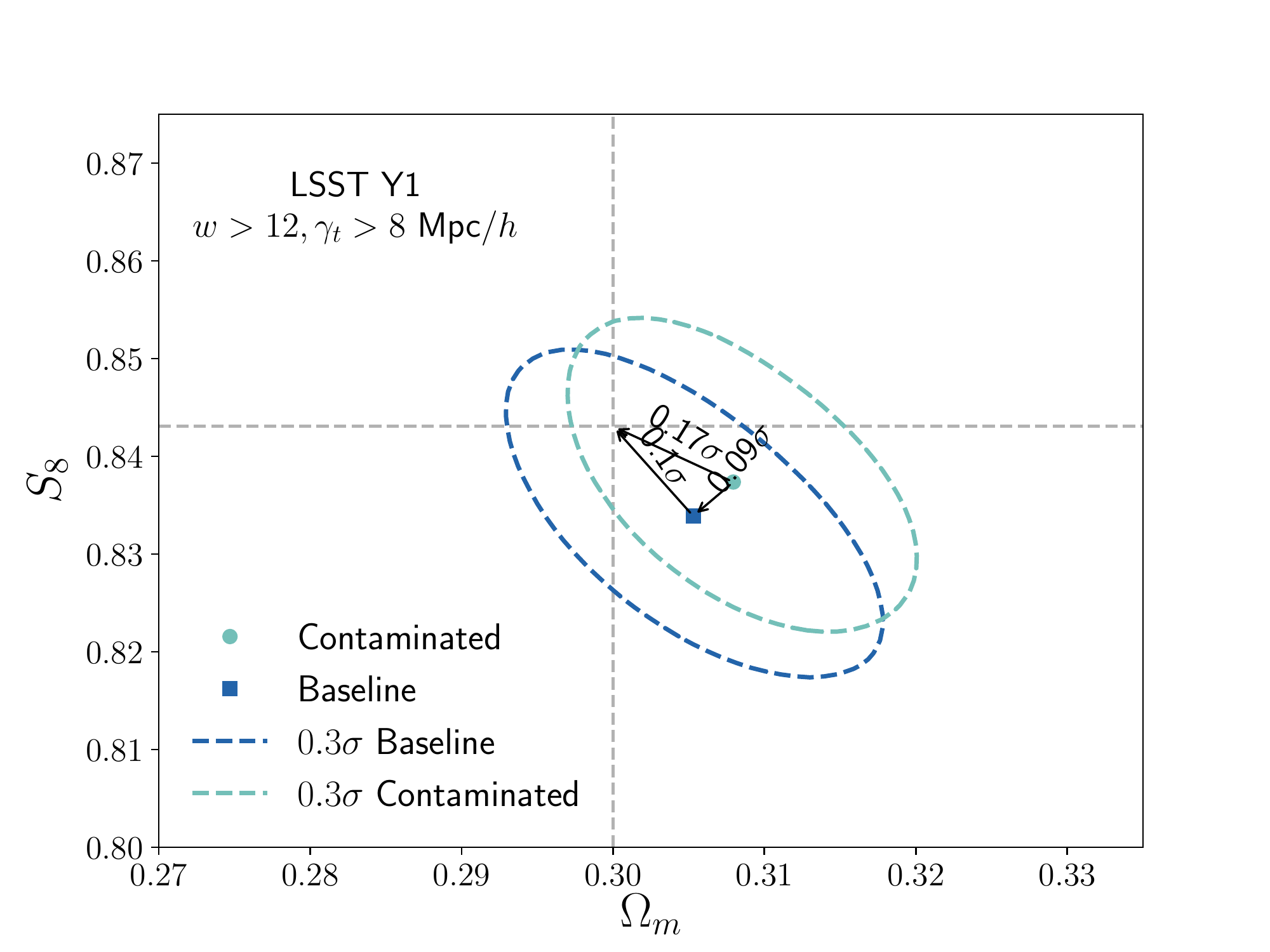}
 \caption{$S_8-\Omega_m$ plane showing the differences in the posteriors using either a fiducial input data vector or a contaminated one with baryonic and non-linear galaxy bias effects. The dashed gray lines mark the input fiducial cosmology. Comparing the contaminated and the baseline posteriors using different sets of scale cuts we have determined that $w>12$ Mpc/$h$, $\gamma_t > 8$ Mpc/$h$ cuts are sufficient for the LSST Y1 setup to recover unbiased results. Specifically, these cuts produce a difference of 0.09 $\sigma$ in the $S_8-\Omega_m$ plane, which is below the threshold of $0.3\sigma$ following the criteria from \citet{y3-methods}}.
\label{fig:scale_cuts}
\end{figure}
In Fig.~\ref{fig:scale_cuts} we show the result of the scale cuts tests for the successful scenario. We have found that scales cuts of $w>12$ Mpc/$h$ and $\gamma_t > 8$ Mpc/$h$ are sufficient in the LSST Y1 setup to recover unbiased results when inputting a contaminated data vector with  non-linear galaxy bias and baryonic effects. We obtain that in this case the difference in the $S_8-\Omega_m$ plane between the fiducial and contaminated data vectors is 0.09$\sigma$ which is below the threshold of $0.3\sigma$, following the same procedure as in \citet{y3-methods}. We compare posteriors using the fiducial vs. the contaminated data vectors (instead of with the input cosmology) to remove any projection effects\footnote{Projection effects are residual differences between the input cosmology and the posteriors under ideal conditions (when the input data vector and model are the same) due to having broad parameter spaces.} impact on this test. We have also tried using  $w>8$ Mpc/$h$, $\gamma_t > 8$ Mpc/$h$ scale cuts, which produced a difference of $0.46\sigma$ and $w>12$ Mpc/$h$ and $\gamma_t > 6$ Mpc/$h$ cuts, which produced a difference of 0.93$\sigma$ in the same $S_8-\Omega_m$ plane. Note that these combinations of scale cuts are an arbitrary choice, and applications on actual data might want to optimize these choices.

\section{Mitigation schemes vs. no mitigation}\label{app:no-mitigation}

\begin{figure}
\begin{center}
\includegraphics[width=0.5\textwidth]{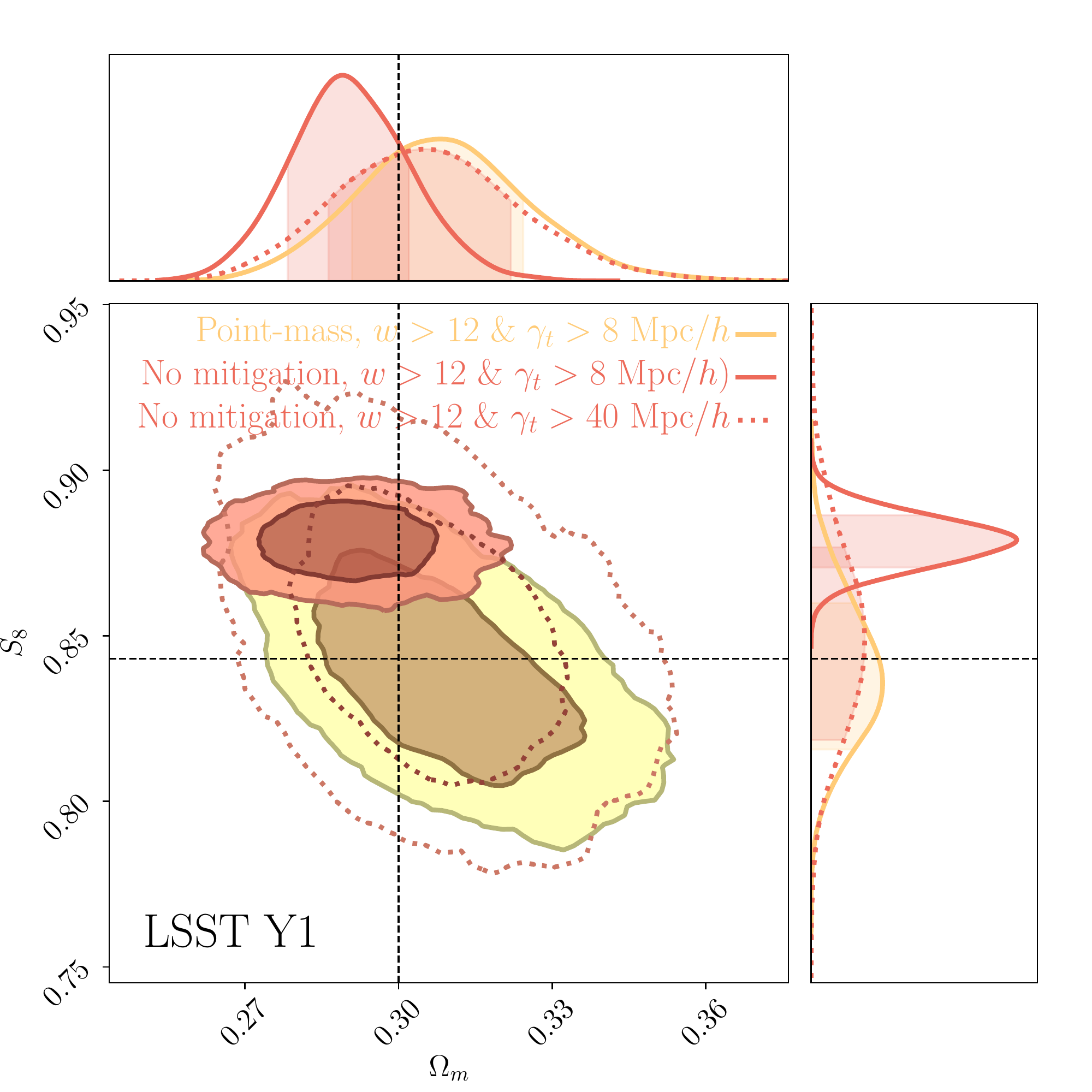}
\caption{We compare the  constraining power when  applying a mitigation scheme (in this case the point-mass marginalization) vs. when not applying any scheme and using less scales for the galaxy-galaxy lensing probe. We conclude that using a marginalization scheme yields $\sim$1.3 more constraining power on the $S_8$ parameter assuming a LSST Y1 simulated scenario. }
\label{fig:no-mitigation}
\end{center}
\end{figure}

In this appendix we address the following question: How much constraining power do we gain by applying one of the mitigation schemes vs. not applying any of them and using fewer scales? To perform  this comparison we choose the point-mass case that includes geometrical information since that is what we used to define the fiducial scale cuts, as described in Sec.~\ref{sec:scale_cuts}.  We perform this comparison for the LSST Y1 simulated analysis. We show the results in Fig.~\ref{fig:no-mitigation}. There we compare the posteriors between using the point-mass mitigation scheme and without applying any mitigation with the following two sets of scale cuts:
\begin{enumerate}
    \item $w>12$ Mpc/$h$, $\gamma_t>8$ Mpc/$h$: the same scale cuts  needed to recover unbiased constraints when applying the point-mass marginalization scheme including geometrical information.
    \item $w>12$ Mpc/$h$, $\gamma_t>40$ Mpc/$h$: scale cuts needed to recover unbiased cosmological constraints without using any mitigation scheme, following the same criteria described in Sec.~\ref{app:scale-cuts}. To obtain them, we have chosen to keep the galaxy clustering scale cut unchanged and increase the tangential shear cut until we recover unbiased results. Under this setup, we find a 0.17$\sigma$ difference in the $S_8-\Omega_m$ plane, while we find a  0.35$\sigma$ difference if we use $\gamma_t$>32 Mpc/$h$ instead, which does not meet the criteria.
\end{enumerate}
We find that $S_8$ is $\sim$1.3 times more constraining when using the point-mass marginalization scheme vs. when not using any mitigation scheme and using larger scale cuts.

\section{Effect on the intrinsic alignment parameters}\label{app:IA}

\begin{figure*}
\begin{center}
\includegraphics[width=0.6\textwidth]{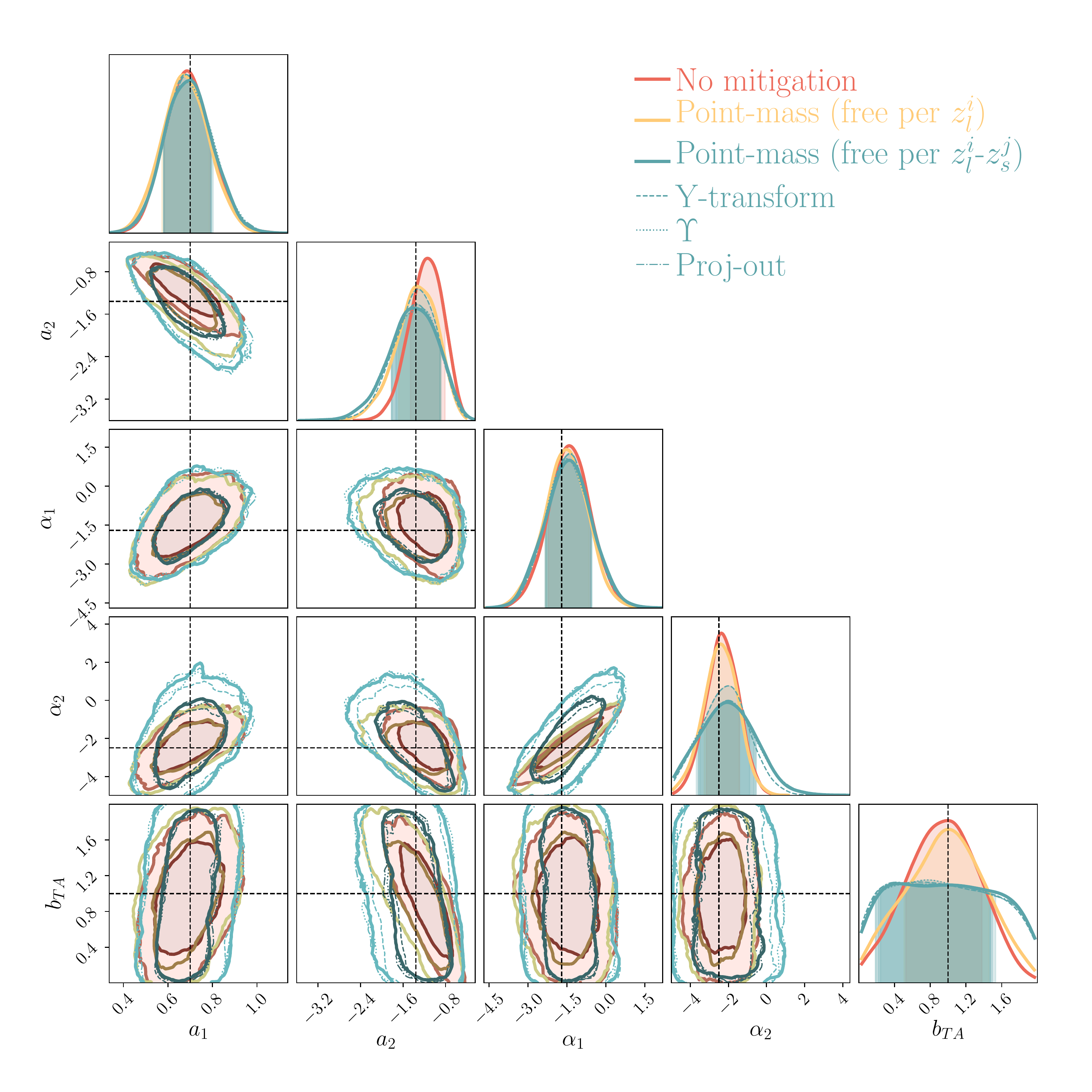}
\includegraphics[width=0.6\textwidth]{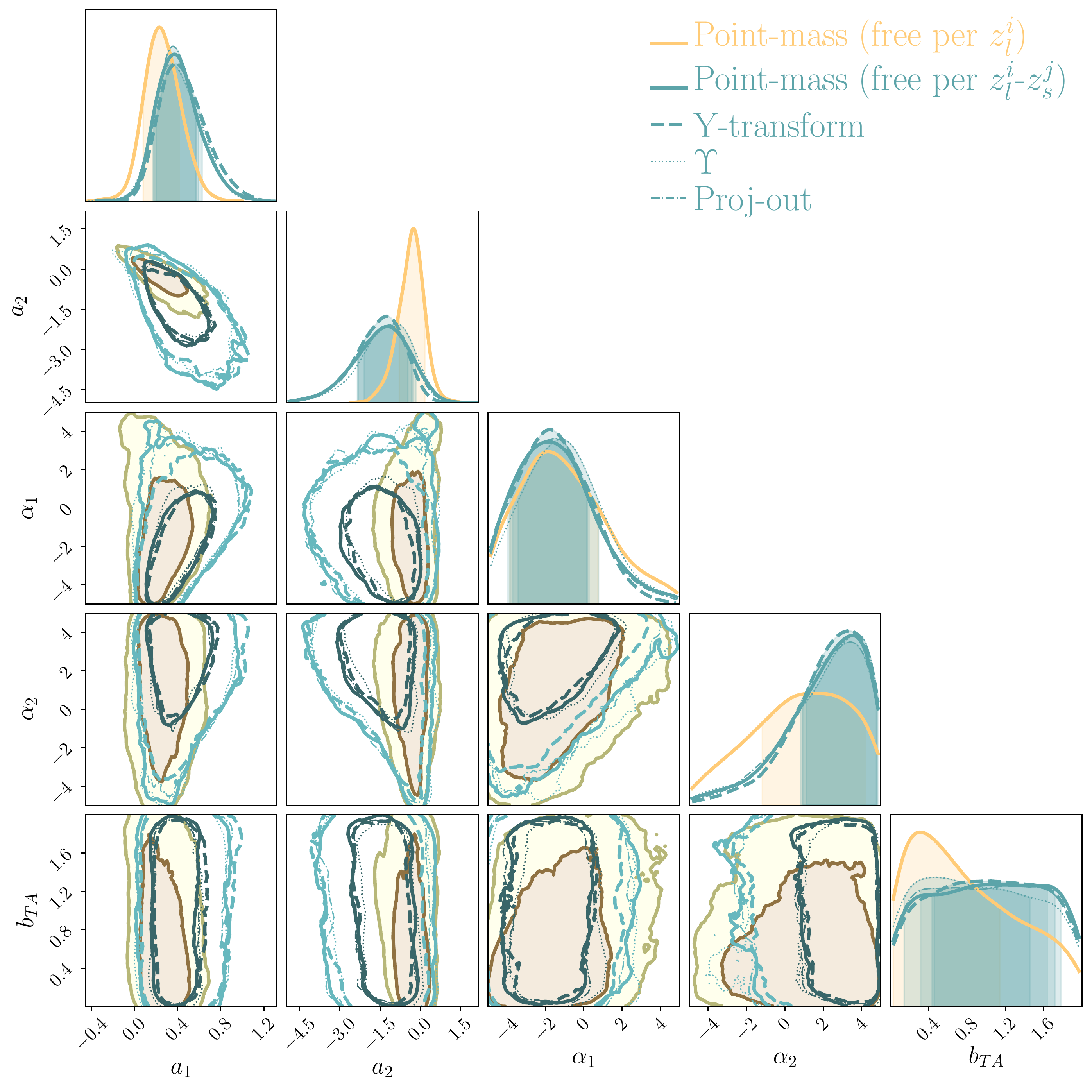}
\caption{\textbf{Top:}  Intrinsic alignment TATT parameter posteriors obtained from an input galaxy clustering and  galaxy-galaxy lensing data vector (2$\times$pt) with non-linear bias and baryonic contamination with a LSST Y1 covariance, comparing the different methodologies to localize the tangential shear measurements. \textbf{Bottom}: Analogous comparison applied to DES Y3 data, for the \textsc{MagLim} 2$\times$pt analysis without the shear-ratio likelihood. }
\label{fig:IA}
\end{center}
\end{figure*}

In Fig.~\ref{fig:IA} we show the posteriors for the tidal alignment and tidal torque (TATT) 5-parameter intrinsic alignment model, at the top for the LSST Y1 simulated case and at the bottom applied to DES Y3 data, in both cases for a 2$\times$2pt analysis without including the shear-ratio likelihood. Using the simulated data in the LSST Y1 setup, we find that using the extra geometric information in the point-mass marginalization, i.e. the case labelled as "Point-mass (free per $z^i_l$)" and described in Sec.~\ref{sec:point-mass-geometric} is beneficial to constrain the intrinsic alignment parameters. In particular we find the parameter describing the galaxy bias of the source sample $b_{TA}$ is more constrained, as well as the parameters affecting the tidal torque part of the model ($a_2$ which describes the amplitude of the IA effect and $\alpha_2$ that modulates its redshift evolution). On the Y3 data, we find that the biggest gain in constraining power is in the $a_2$ parameter.

\section{Comparison between DES Y3 and LSST Y1}\label{app:desy3vslssty1}

\begin{figure*}
\begin{center}
\includegraphics[width=0.7\textwidth]{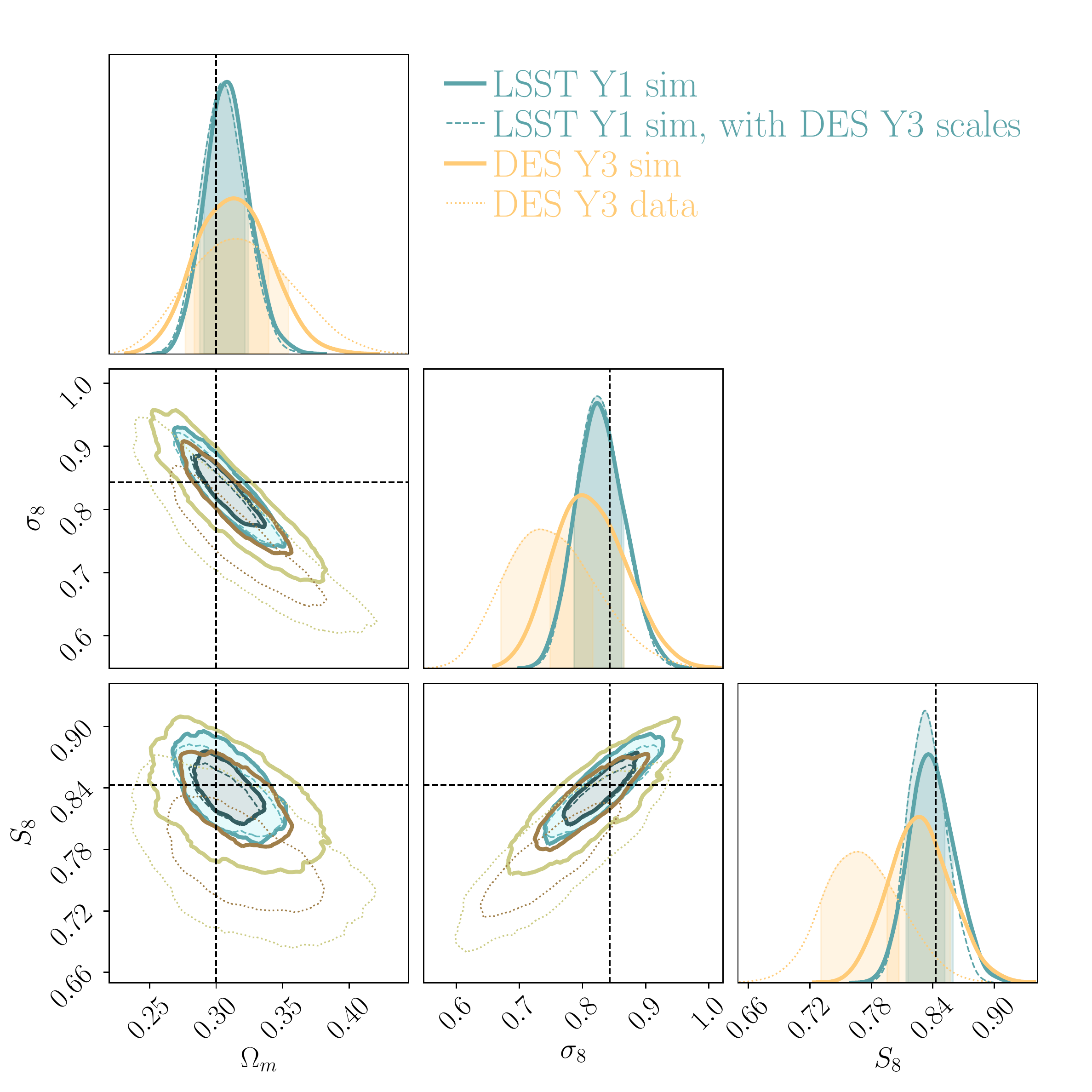}
\end{center}
\caption{Comparison of the constraining power between LSST Y1 and DES Y3 for a 2$\times$2pt simulated analysis using the fiducial model data vectors. For LSST Y1 we use scales $w>12$ Mpc/$h$, $\gamma_t > 8$ Mpc/$h$, and for DES Y3  $w>8$ Mpc/$h$, $\gamma_t > 6$ Mpc/$h$. We also include the data results from DES Y3 2$\times$2pt.}
\label{fig:y3_vs_lsst}
\end{figure*}

In Fig.~\ref{fig:y3_vs_lsst} we compare the constraining power between the  DES Y3 and LSST Y1 setups, which besides being interesting on its own also provides some basic  validation of the LSST Y1 covariance that we compute with \textsc{CosmoCOV}. First, we compare the fiducial LSST Y1 simulated analysis with the fiducial scales of $w>12$ Mpc/$h$ and $\gamma_t > 8$ Mpc/$h$, with the scales used in the DES Y3 data of $w>8$ Mpc/$h$ and $\gamma_t > 6$ Mpc/$h$. We observe that the degradation in the constraining power coming only from the differences in the scales is quite small. Then, we compare the size of the  LSST Y1 contours with the simulated DES Y3 analysis, which was used in \citet{y3-methods} to determine the scale cuts for that case. The only two differences between the contours labelled "LSST Y1 sim, with DES Y3 scales" and the ones labelled "DES Y3 sim" are the covariances and the input redshift distributions, as described in Sec.~\ref{sec:fiducialmodel}. Finally we also compare the DES Y3 simulated analysis with the actual final data DES Y3 2$\times$2pt results, which use a different set of priors as the rest of the chains, had an updated covariance accounting for the best-fit parameters and are not centered at the same cosmology.

\end{document}